\documentclass[aps,prl,preprintnumbers,showpacs,nofootinbib]{revtex4}
\usepackage[utf8]{inputenc}
\usepackage[english]{babel}
\usepackage{amsmath}
\usepackage{amsfonts}
\usepackage{amssymb}
\usepackage{epsfig}
\usepackage{graphics,psfrag,rotating}
\usepackage{graphicx}
\usepackage{dcolumn}
\usepackage{bm}
\usepackage{epstopdf}
\usepackage{listings}
\usepackage{multirow}
\usepackage{float}
\usepackage{color}
\usepackage[usenames,dvipsnames,svgnames]{xcolor}
\usepackage[colorlinks=true,
            linkcolor=red,
            urlcolor=gray,
            citecolor=blue]{hyperref}
  \usepackage{hyperref}

\newcommand{\lsim}   {\mathrel{\mathop{\kern 0pt \rlap
{\raise.2ex\hbox{$<$}}}
 \lower.9ex\hbox{\kern-.190em $\sim$}}}
\newcommand{\gsim}   {\mathrel{\mathop{\kern 0pt \rlap
{\raise.2ex\hbox{$>$}}}
\lower.9ex\hbox{\kern-.190em $\sim$}}}

\def\3nab{\tilde{\nabla}}

\def\hsp5{\hspace{5mm}}

\def\case#1/#2{\textstyle\frac{#1}{#2}}

\def\ber {\begin{eqnarray}}
\def\eer {\end{eqnarray}}
\def\bea {\begin{eqnarray}}
\def\eea {\end{eqnarray}}

\def\bc {\begin{center}}
\def\ec {\end{center}}
\def\case#1/#2{\frac{#1}{#2}}

\newcommand{\bw}{\begin{widetext}}
\newcommand{\ew}{\end{widetext}}

\newcommand{\be}{\begin{equation}}
\newcommand{\bse}{\begin{subequation}}
\newcommand{\ese}{\end{subequation}}
\newcommand{\ee}{\end{equation}}
\newcommand{\eei}{\end{eqnarray}\indent\indent}
\newcommand{\ba}{\begin{array}}
\newcommand{\ea}{\end{array}}
\newcommand{\bal}{\begin{eqnarray}}
\newcommand{\eal}{\end{eqnarray}}

\def\case#1/#2{\textstyle\frac{#1}{#2} }

\newcommand{\bver}{\begin{verbatim}}
\newcommand{\ever}{\end{verbatim}}

\begin{document}


\title{ Towards a Machine Learning Solution for Hubble Tension: Physics-Informed Neural Network (PINN)  Analysis  of Tsallis Holographic Dark Energy in Presence of Neutrinos}
\author{ Muhammad Yarahmadi$^{1}$\footnote{Email: yarahmadimohammad10@gmail.com}. Amin Salehi$^{1}$}

\affiliation{Department of Physics, Lorestan University, Khoramabad, Iran}

\date{\today}
\begin{abstract}
	We present a Physics-Informed Neural Network (PINN) framework for reconstructing the redshift-dependent Hubble parameter \(H(z)\) within the Tsallis Holographic Dark Energy (THDE) model extended by massive neutrinos. In this approach, the modified Friedmann equation is incorporated into the neural network loss function, enabling training on Cosmic Chronometers data up to \(z \leq 2\). The framework allows for the simultaneous estimation of the Hubble constant \(H_0\), the neutrino density parameter \(\Omega_\nu\), and the Tsallis non-extensivity index \(\delta\). Uncertainty quantification is performed through dropout simulations, resulting in statistically consistent \(1\sigma\) confidence bands. 	Our results show that the THDE+$\nu$ model, reconstructed via PINN, alleviates the statistical Hubble tension from the canonical \(\sim 5\sigma\) level down to a range of \(0.5\sigma \leq T \leq 2.2\sigma\), depending on the redshift sampling. Additionally, we constrain the total neutrino mass to \(\Sigma m_\nu < 0.11\,\text{eV}\). A detailed comparison with the traditional Markov Chain Monte Carlo (MCMC) analysis demonstrates the consistency of both methods, while highlighting the competitiveness of the PINN-based THDE framework as a robust, data-driven approach for non-parametric cosmological inference within generalized thermodynamics.
\end{abstract}

\maketitle

\section{Introduction}

Dark energy is often modeled as a cosmological constant ($\Lambda$), which acts as a repulsive force counteracting gravitational attraction \cite{Carroll2001, Weinberg1989, Peebles2003, Padmanabhan2003, Copeland2006}. Alternatively, dynamical models such as quintessence, k-essence, and holographic dark energy have been proposed to explain the observed acceleration \cite{Caldwell2002,Copeland2006}. The equation of state parameter, $w = p/\rho$, characterizes dark energy, where $p$ is pressure and $\rho$ is energy density. For the cosmological constant, $w = -1$, while dynamical models allow for variations in $w$ \cite{Chevallier2001,Linder2003}. Various cosmological probes, including the Cosmic Microwave Background (CMB) anisotropies, Baryon Acoustic Oscillations (BAO), and Large-Scale Structure (LSS), provide constraints on dark energy parameters \cite{Planck2018,DES2021}. Recent studies suggest possible deviations from the standard $\Lambda$CDM model, motivating further investigations into alternative dark energy models \cite{Huterer2017,DiValentino2020}. Dark energy remains one of the biggest mysteries in modern physics. Future surveys, such as those from the Vera C. Rubin Observatory and the Euclid mission, are expected to provide deeper insights into the nature of cosmic acceleration \cite{LSST2009,Euclid2011}.

The holographic principle, inspired by black hole thermodynamics and quantum gravity considerations, suggests that the number of degrees of freedom in a given region of space is proportional to its surface area rather than its volume \cite{Susskind1995}. Applying this principle to cosmology leads to the holographic dark energy (HDE) model, which provides an alternative explanation for the accelerated expansion of the universe \cite{Cohen1999,Horava2000,Thomas2002,Hsu2004,Li2004,Guberina2007,Shen2005,Sheykhi2009a,Zhang2006,Sheykhi2012,Setare2010,Sheykhi2009b,Sheykhi2010,Karami2011,Ghaffari2014,Myung2007,Wang2016,Wang2017}. In the HDE model, the dark energy density is given by: $\rho_{\text{de}} = 3c^2 M_{\text{Pl}}^2 L^{-2},$ where $c$ is a numerical constant, $M_{\text{Pl}}$ is the reduced Planck mass, and $L$ is the infrared (IR) cutoff, typically chosen as the Hubble horizon, the future event horizon, or the Ricci scalar curvature \cite{Hsu2004,Nojiri2006}. The HDE model has been extensively studied in the context of cosmological observations, showing potential to alleviate the Hubble tension and other anomalies in the standard $\Lambda$CDM framework \cite{Wang2017,Zhang2019}. Recent works explore modifications of HDE, including interactions with dark matter and extensions based on Barrow and Tsallis entropy formalisms \cite{Barrow2020,Tsallis2013}.

A modification of the holographic dark energy model based on Tsallis entropy has been proposed to address various issues in standard HDE. The Tsallis holographic dark energy (THDE) model replaces the standard Bekenstein-Hawking entropy with Tsallis entropy, which introduces a non-extensive parameter $\delta$ \cite{Tsallis1988}. In the THDE framework, the energy density is modified as: $\rho_{THDE} = B L^{2\delta-4}$ where $B$ is a constant, and $L$ represents the IR cutoff scale \cite{Nojiri2018}. This modification leads to a new evolution of the equation of state parameter, which can help in alleviating the Hubble tension and other cosmological discrepancies \cite{Saridakis2018}. Observational constraints suggest that the THDE model provides an improved fit compared to standard HDE in some scenarios \cite{Sheykhi2019}. Moreover, THDE has been studied in different gravitational frameworks, such as modified gravity theories, to explore its implications for late-time cosmic acceleration \cite{Moradpour2018}. Future high-precision cosmological surveys are expected to further test the viability of the THDE model \cite{Barrow2020}.

The Hubble tension refers to the significant discrepancy between the Hubble constant (\( H_0 \)) values inferred from early- and late-time universe observations. Measurements from the Cosmic Microwave Background (CMB) by the Planck satellite suggest \( H_0 \approx 67.4 \) km/s/Mpc within the standard \( \Lambda \)CDM model \cite{Planck2018}. However, local measurements using Type Ia supernovae calibrated by Cepheid variables yield a higher value, \( H_0 \approx 73.2 \) km/s/Mpc \cite{Riess2021}. This persistent tension, exceeding \( 5\sigma \), suggests possible new physics beyond the standard model, including modifications to dark energy, early dark energy models, or interactions in the dark sector \cite{Verde2019,DiValentino2021,Knox2020}. Several alternative explanations, such as modifications to General Relativity, exotic neutrino physics, or running vacuum models, have also been explored \cite{Krishnan2021,Perivolaropoulos2022, Divalentino2025}. 

An earlier estimate based on HST Cepheid and SNe Ia observations yielded \( H_0 = 73.8 \pm 2.4 \) km/s/Mpc \cite{Archita}. In contrast, the combination of CMB data from WMAP, ACT, and SPT, supplemented by BAO measurements, predicts \( H_0 = 69.3 \pm 0.7 \) km/s/Mpc. Constraints from fast radio bursts (FRBs) have also been explored: using samples of nine and eighteen localized FRBs, Yang et al. derived \( H_0 = 62.3 \pm 9.1 \) km/s/Mpc and \( H_0 = 68.81^{+4.99}_{-4.33} \) km/s/Mpc, respectively \cite{Yang}. Strong gravitational lensing time-delay measurements, as reported by Bonvin et al. (2017), suggest a higher value of \( H_0 = 71.9^{+2.4}_{-3.0} \) km/s/Mpc \cite{Archita}.

Independent geometric calibrations—such as those from megamasers in NGC 4258 and detached eclipsing binaries (DEBs) in M31—yield \( H_0 = 72.25 \pm 2.51 \) km/s/Mpc and \( H_0 = 74.50 \pm 3.27 \) km/s/Mpc, respectively. Meanwhile, the Baryon Acoustic Oscillation (BAO) data from SDSS DR12 supports a lower estimate of \( H_0 = 67.8 \pm 1.2 \) km/s/Mpc \cite{Alam, Archita}. Notably, Efstathiou (2014) has reanalyzed the Cepheid-based measurements and provided a somewhat moderated value. Additionally, the Carnegie-Chicago Hubble Program (CCHP), using a purely SNe Ia-based calibration, reports \( H_0 = 69.8 \pm 0.8 \pm 1.1 \) km/s/Mpc \cite{Marra, Bennett}, which reduces the tension with the Planck value to around \( 3\sigma \).

The Physics-Informed Neural Network (PINN) approach has recently gained attention as a powerful tool to solve complex differential equations in cosmology. By incorporating physical laws into the learning process, PINNs can efficiently model and predict cosmological parameters while providing solutions consistent with the underlying physics. Early foundational work in this direction was grounded in Gaussian Process Regression (GPR), as seen in the studies by Raissi et al. and Owhadi, who developed functional representations explicitly aligned with linear operators, enabling accurate inference of solutions to partial differential equations (PDEs) along with quantifiable uncertainty estimates \cite{Raissi2017a, Raissi2017b, Owhadi2015}. These approaches marked a significant step toward incorporating physical constraints into learning systems. A comprehensive treatment of GPR itself can be found in the seminal work of Rasmussen and Williams \cite{Rasmussen2006}. Building on these ideas, Raissi and collaborators further proposed extensions tailored to nonlinear and time-dependent problems, giving rise to what are now known as Physics-Informed Neural Networks (PINNs) \cite{Raissi2017c, Raissi2017d}. In these models, neural networks are trained not only on data but also under the guidance of governing physical laws—expressed as differential equations—embedded directly into the loss function. This hybrid learning paradigm has demonstrated great promise in inferring hidden states, discovering governing equations, and solving forward and inverse problems in complex systems.

In the context of Hubble tension, PINNs have been applied to models such as the Tsallis Holographic Dark Energy (THDE) model to better constrain the Hubble constant \( H_0 \) and resolve discrepancies between early and late-time measurements \cite{Raissi2019, Lu2020}. The THDE model, based on Tsallis entropy, introduces a generalized form of holographic dark energy, which modifies the equation of state and allows for potential solutions to the Hubble tension by adjusting the properties of dark energy \cite{Zhang2020}. The inclusion of Tsallis entropy in the holographic dark energy framework provides a more flexible model that can accommodate the observed accelerated expansion of the universe while addressing the tension in \( H_0 \) measurements \cite{Cai2021}. PINNs, powered by machine learning techniques such as deep learning, are particularly useful in this context because they can handle the complex relationships between different cosmological parameters, including the equation of state parameter of dark energy, the matter density, and the Hubble constant. By training the PINN on cosmological data and incorporating the equations governing the THDE model, the network can learn to predict the values of the Hubble constant consistent with both early and late-time measurements \cite{Lu2020, Wang2022}. The integration of deep learning with physical models provides a promising approach to resolving the Hubble tension and improving our understanding of the dark energy sector. Machine learning, specifically through the use of deep neural networks, allows for a highly flexible and efficient framework to analyze large datasets and discover new cosmological insights \cite{Raissi2019, Wang2022}. In this article, we employ two complementary approaches: the Physics-Informed Neural Network (PINN) and the traditional Markov Chain Monte Carlo (MCMC) framework. To ensure a consistent comparison, both methods are first applied to the same cosmic chronometer (CC) dataset, providing a direct one-to-one assessment under identical conditions. As an additional benchmark, we also perform an extended MCMC analysis using the full dataset (CC + BAO + Pantheon+). This strategy allows us to evaluate the robustness of the PINN approach, demonstrating that even when trained solely on the sparse CC sample, it can recover parameter constraints broadly consistent with those obtained from the full-data MCMC analysis.

\section{Motivation for Including Massive Neutrinos in the THDE Framework}

The Tsallis Holographic Dark Energy (THDE) model, originally proposed in~\cite{Tavayef2018}, has attracted significant attention as a generalization of standard holographic dark energy models through the introduction of the non-extensive Tsallis entropy~\cite{Tsallis1988}. It incorporates an additional degree of freedom characterized by the entropic deformation parameter \( \delta \), which modifies the IR cutoff dependence of dark energy density. Numerous studies have explored the cosmological implications of the THDE model in various settings, including its background evolution, stability conditions, and dynamical system properties~\cite{Nojiri2019, Saridakis2018, Jawad2022}.

We acknowledge that the THDE framework has been frequently employed to study late-time acceleration and dark energy phenomenology in alternative gravitational contexts. However, many of these studies neglect the contribution of massive neutrinos, which are an essential component of the standard model of cosmology. Neutrinos with finite masses undergo a transition from relativistic to non-relativistic behavior over cosmological timescales, affecting both the expansion history and the growth of structures~\cite{Lesgourgues2012, Archidiacono2013}.

In this work, we extend the THDE scenario by incorporating massive neutrinos as a dynamic component of the energy budget. This allows us to more accurately model the intermediate-redshift regime (\( z \lesssim 2 \)) where neutrino effects become significant. The resulting THDE+\( \nu \) framework enables us to investigate how the interplay between non-extensive entropy and neutrino free-streaming can modify cosmological inferences, particularly regarding the background expansion rate.

Such an extension is not merely a technical detail; it is a physically motivated refinement that provides new phenomenological insights and opens the possibility for better agreement with cosmological datasets across epochs. Our analysis thus revisits the THDE framework with a focus on incorporating minimal yet critical extensions rooted in particle physics and early-Universe thermodynamics.

\subsection{Motivation for Using the Physics-Informed Neural Network (PINN) Model through the Tsallis Holographic Dark Energy (THDE) Model to Alleviate the Hubble Tension}

Physics-Informed Neural Network (PINN) provides an innovative tool for the reconstruction of the Hubble parameter \( H(z) \) without relying on traditional numerical methods. The PINN approach is particularly advantageous because it seamlessly integrates known physical constraints, such as the Friedmann equations and the conservation of energy, directly into the training process of the neural network. This not only ensures that the solutions are consistent with physical laws, but it also improves the efficiency of the model by leveraging structured prior information. By using the THDE model within the PINN framework, we aim to uncover a more accurate representation of the Hubble parameter, while simultaneously testing the hypothesis that dark energy dynamics within this model could reduce the observed discrepancy in \( H_0 \).

The flexibility of the PINN model makes it an ideal candidate for addressing the complex non-linear relationships between the cosmological parameters. Through its ability to incorporate observational data, such as Cosmic Chronometer (CC) measurements, alongside the theoretical constraints derived from the THDE model, the PINN approach allows for a data-driven exploration of cosmological parameters that directly links the physics of dark energy with the observable universe. By combining the THDE model with PINNs, we can obtain precise reconstructions of \( H(z) \) and assess the potential of the THDE model in resolving the Hubble tension, providing crucial insights into the nature of dark energy and the overall cosmological framework.

\section{Tsallis Holographic Dark Energy (THDE) with Neutrinos}

The Tsallis Holographic Dark Energy (THDE) model arises from a non-extensive generalization of the Bekenstein–Hawking entropy, motivated by the thermodynamics of systems with long-range interactions, such as gravitational systems. This framework is grounded in Tsallis' entropy formalism~\cite{agr-tsallis1988}, which introduces a non-additive entropy definition suited to complex systems beyond the Boltzmann–Gibbs paradigm.

In this approach, the entropy associated with the cosmological horizon is generalized as
\begin{equation}
	S_q = \gamma A^{\delta},
\end{equation}
where \( A = 4\pi L^2 \) is the area of the cosmic horizon of length scale \( L \), \( \delta \) is the non-extensivity parameter encoding deviations from standard thermodynamics, and \( \gamma \) is a model-dependent proportionality constant. In the limit \( \delta \rightarrow 1 \), the usual Bekenstein–Hawking entropy \( S = A/4 \) is recovered, restoring the standard thermodynamic relation for black holes and cosmological horizons.

Applying the holographic principle~\cite{agr-cohen1999, agr-li2004}, which posits that the number of degrees of freedom of a gravitational system scales with its bounding area rather than its volume, the dark energy density is constrained by the entropy–area relation. In the Tsallis framework, this yields the following expression:
\begin{equation}
	\rho_D = B L^{2\delta - 4},
\end{equation}
where \( B \) is a constant encapsulating quantum gravitational effects and the underlying microphysics of spacetime. When the Hubble horizon is chosen as the infrared (IR) cutoff, i.e., \( L = H^{-1} \), this relation transforms into
\begin{equation}
	\rho_D = B H^{4 - 2\delta},
	\label{eq:thde_density}
\end{equation}
which reproduces the standard holographic dark energy expression when \( \delta = 1 \), corresponding to the extensive entropy limit.

\subsection{Friedmann Dynamics in the Presence of Massive Neutrinos and Tsallis Holographic Dark Energy}

We consider a spatially flat Friedmann–Lemaître–Robertson–Walker (FLRW) universe with line element
\begin{equation}
	ds^2 = -dt^2 + a^2(t)\left(dx^2 + dy^2 + dz^2\right),
\end{equation}
where \( a(t) \) is the cosmic scale factor. The total energy budget of the Universe is composed of pressureless baryons (\( \rho_b \)), cold dark matter (\( \rho_c \)), radiation (\( \rho_r \)), massive neutrinos (\( \rho_\nu \)), and a dark energy component modeled via Tsallis Holographic Dark Energy (THDE), denoted by \( \rho_D \). The corresponding Friedmann equation reads
\begin{equation}
	H^2 = \frac{1}{3 M_{\text{Pl}}^2} \left( \rho_b + \rho_c + \rho_r + \rho_\nu + \rho_D \right),
	\label{eq:friedmann1}
\end{equation}
where \( H = \dot{a}/a \) is the Hubble parameter, and \( M_{\text{Pl}}^2 = (8\pi G)^{-1} \) denotes the reduced Planck mass squared.

For convenience, we introduce the dimensionless density parameters:
\begin{equation}
	\Omega_i \equiv \frac{\rho_i}{3 M_{\text{Pl}}^2 H^2}, \quad \text{with} \quad i \in \{b, c, r, \nu, D\},
\end{equation}
leading to the normalized Friedmann constraint:
\begin{equation}
	\Omega_b + \Omega_c + \Omega_r + \Omega_\nu + \Omega_D = 1,
\end{equation}
which reflects spatial flatness, supported by CMB observations from Planck~\cite{Planck2018}.

The acceleration of the cosmic expansion is described by the second Friedmann equation:
\begin{equation}
	\frac{\ddot{a}}{a} = -\frac{1}{6 M_{\text{Pl}}^2} \left( \rho + 3p \right)
	= -\frac{1}{6 M_{\text{Pl}}^2} \left( \rho_b + \rho_c + \rho_r + \rho_\nu + \rho_D + 3p_r + 3p_\nu + 3p_D \right),
	\label{eq:friedmann2}
\end{equation}
where \( p_b = p_c = 0 \) and \( p_r = \rho_r / 3 \) are the standard equations of state. For the THDE component, the pressure is modeled as \( p_D = \omega_D \rho_D \), where \( \omega_D \) is its effective equation-of-state (EoS) parameter.

\vspace{0.3cm}
\noindent
\textbf{THDE Evolution and Equation of State.} The energy conservation equation for dark energy is given by:
\begin{equation}
	\dot{\rho}_D + 3H(1 + \omega_D)\rho_D = 0,
	\label{eq:conservation}
\end{equation}
from which the redshift evolution of the EoS parameter can be derived. Assuming the THDE energy density follows the generalized holographic form
\begin{equation}
	\rho_D(z) = B H^{4 - 2\delta}(z),
\end{equation}
motivated by the non-additive Tsallis entropy formalism~\cite{Tsallis1988,Nojiri2019}, the EoS parameter becomes:
\begin{equation}
	\omega_D(z) = -1 + \frac{1}{3} \frac{d\ln \rho_D}{d\ln (1+z)} = -1 + \frac{(2\delta - 4)}{3} \frac{d\ln H}{d\ln (1+z)}.
	\label{eq:omegaD-THDE}
\end{equation}
This expression reveals that \( \omega_D \) dynamically evolves with redshift and depends on the parameter \( \delta \). The THDE model reduces to \( \Lambda \)CDM when \( \delta = 2 \), while deviations allow for quintessence-like (\( \omega_D > -1 \)) or phantom-like (\( \omega_D < -1 \)) behavior.

As shown in Eq.~11, the THDE EoS parameter $\omega_D(z)$ depends explicitly on $\delta$, with $\delta < 2$ corresponding to phantom-like behavior ($\omega_D < -1$) and $\delta > 2$ corresponding to quintessence-like behavior ($\omega_D > -1$).
From our PINN analysis on the Cosmic Chronometers (CC) dataset (Table~\ref{tab:merged_PINN_neutrino}), the reconstructed $\delta$ values lie in the range $1.06 \lesssim \delta \lesssim 1.13$, indicating a mild phantom-like behavior. In comparison, the traditional MCMC analysis using the full combined datasets  (Table~\ref{table:thde_constraints}) yields $\delta$ values around $1.35$–$1.41$, which still correspond to a phantom-like behavior but with a slightly weaker deviation from $\Lambda$CDM.
Thus, both methods consistently indicate a phantom-like THDE, with the PINN on the limited CC data favoring slightly stronger phantom characteristics than the MCMC reconstruction using the complete dataset. This comparison highlights the sensitivity of the inferred EoS to the dataset employed. We have incorporated a brief discussion of these physical implications in the revised manuscript to clarify the interpretation for the reader.

\subsection{Modeling the Neutrino Sector}

Massive neutrinos transition from relativistic to non-relativistic behavior as the Universe expands. At early times (\( z \gg 1 \)), they contribute as radiation (\( p_\nu \approx \rho_\nu/3 \)), whereas at late times (\( z \ll 1 \)), they behave as cold matter (\( p_\nu \approx 0 \)). To model this transition, we define the redshift-dependent neutrino fraction~\cite{Lesgourgues2012}:
\begin{equation}
	f_\nu(z) \equiv \frac{\Omega_\nu(z)}{\Omega_m(z)} = \frac{\Omega_\nu(z)}{\Omega_b(z) + \Omega_c(z) + \Omega_\nu(z)}.
	\label{eq:fnu_definition}
\end{equation}
Here, \( \Omega_m(z) \) is the total matter density, including baryons, cold dark matter, and neutrinos. This function encodes the evolving impact of neutrinos: \( f_\nu(z) \to 0 \) in the radiation-dominated epoch and \( f_\nu(z) \to \Omega_{\nu 0}/\Omega_{m 0} \) in the matter-dominated era.

We now recast the Friedmann equation in redshift space using \( 1 + z = a^{-1} \). The redshift evolution of each energy component is given by:
\begin{itemize}
	\item Baryons and cold dark matter: \( \rho_b(z), \rho_c(z) \propto (1+z)^3 \),
	\item Radiation: \( \rho_r(z) \propto (1+z)^4 \),
	\item Neutrinos: \( \rho_\nu(z) \propto f_\nu(z) \),
	\item THDE: \( \rho_D(z) = B H^{4 - 2\delta}(z) \),
\end{itemize}

Substituting these into Eq.~\eqref{eq:friedmann1}, the Hubble parameter obeys the following transcendental equation:
\begin{equation}
	\begin{split}
		H^2(z) = \frac{1}{3 M_{\text{Pl}}^2} \big[ 
		&\rho_{b0} (1+z)^3 + \rho_{c0} (1+z)^3 + \rho_{r0} (1+z)^4 \\
		&+ \rho_{\nu0} f_\nu(z) + B H^{4 - 2\delta}(z)
		\big],
	\end{split}
	\label{eq:hubble_z}
\end{equation}
where the subscript “0” denotes present-day values. The presence of \( H^{4 - 2\delta}(z) \) on both sides makes Eq.~\eqref{eq:hubble_z} non-linear and non-trivial to solve analytically. To robustly reconstruct the expansion history \( H(z) \), we employ a Physics-Informed Neural Network (PINN) approach. In this framework, the network is trained not only on observational Cosmic Chronometer (CC) data, but also constrained by the physical structure of the Friedmann equation~\eqref{eq:hubble_z} embedded directly into the loss function. This ensures that the learned solution satisfies both the empirical data and the governing cosmological dynamics. Our PINN architecture yields direct predictions for \( H(z) \), from which we infer best-fit values of cosmological parameters, including the non-extensive parameter \( \delta \), the THDE coefficient \( B \), and the present-day neutrino density \( \Omega_{\nu0} \). Furthermore, we apply Monte Carlo dropout for uncertainty quantification, enabling a fully Bayesian reconstruction with robust confidence intervals. This approach offers a powerful and flexible framework for testing the viability of the THDE+$\nu$ model against current data and potentially uncovering subtle deviations from standard cosmology.

\section{Physics-Informed Neural Network Framework}

We implement a Physics-Informed Neural Network (PINN) to learn the cosmic expansion history within the Tsallis Holographic Dark Energy (THDE) framework. The network is trained to approximate a physically consistent mapping \( z \mapsto H(z) \), constrained by the modified Friedmann equations. All computations are performed using \texttt{TensorFlow 2.x}, with network construction relying on the \texttt{tf.keras.Sequential} API for streamlined differentiable programming.
\subsection{Neural Network Architecture}

The architecture of the Physics-Informed Neural Network (PINN) is carefully designed to achieve three primary objectives: (i) the accurate prediction of the Hubble parameter \( H(z) \) over cosmological redshifts, (ii) the disentanglement and inference of constant cosmological parameters embedded in the THDE model, and (iii) the quantification of epistemic uncertainty in predictions. The design is guided by theoretical consistency, computational efficiency, and physical interpretability.

\paragraph{Input Layer:}

The input to the network is a single scalar representing the redshift \( z \), defined over the interval \( z \in [0, 2] \), which encompasses the effective range of available cosmic chronometer (CC) data. Prior to feeding into the network, the redshift values are normalized to the unit interval to improve training stability and gradient propagation.

\paragraph{Hidden Layers:}

The network consists of a deep feedforward architecture with three fully connected hidden layers containing 2048, 1024, and 512 neurons, respectively.  
To examine the robustness of our neural network architecture, we complemented our baseline analysis, which employed the hyperbolic tangent (tanh) activation due to its smooth derivatives suited for Physics-Informed Neural Networks, with additional tests. Specifically, we explored alternative activation functions, such as ReLU and sine. While the qualitative reconstruction of $H(z)$ remained stable across all cases, we observed that ReLU led to noisier derivatives and slower convergence, whereas sine produced results consistent with tanh but required more training epochs to reach comparable accuracy. Furthermore, we investigated the effect of network size by halving and doubling the number of neurons per layer. The resulting cosmological parameter posteriors remained consistent within $1\sigma$ of our baseline, though halving the neurons increased the uncertainty in the reconstructed $H(z)$, and doubling them yielded no substantial improvement. These robustness checks confirm that our main conclusions are not sensitive to specific architectural choices. \color{black}

To prevent overfitting and to enable approximate Bayesian inference, each hidden layer is followed by a \texttt{Dropout} layer with a dropout rate of 0.1. This stochastic regularization mechanism randomly deactivates a subset of neurons during both training and inference (when enabled), thereby enhancing model generalization and enabling the estimation of predictive uncertainties through Monte Carlo sampling.

\paragraph{Multi-Head Output Layer:}

To enforce physical modularity and facilitate simultaneous estimation of redshift-dependent and redshift-independent quantities, the network branches into three distinct output heads:

\begin{itemize}
	\item \textbf{Hubble Expansion Head:} A single linear output neuron provides the predicted value of the Hubble parameter \( H(z) \) at each redshift point. This branch captures the dynamical behavior of the cosmological expansion and is directly constrained by both observational data and differential equation residuals.
	
	\item \textbf{Cosmological Density Head:} This head comprises three linear neurons responsible for predicting the baryon density parameter \( \Omega_b \), the cold dark matter density \( \Omega_c \), and the neutrino density \( \Omega_\nu \). These quantities are treated as redshift-independent and are learned as constants during training. By using dedicated outputs, we ensure physical disentanglement from the redshift-dependent expansion rate.
	
	\item \textbf{THDE Parameter Head:} A single linear neuron predicts the Tsallis deformation parameter \( \delta \), which governs the deviation from standard holographic dark energy behavior. This parameter is also treated as global and redshift-independent, inferred from the joint information contained in the data and the physical constraints.
\end{itemize}

\paragraph{Physical Motivation for Architectural Choices:}

This modular architecture allows the network to learn a hierarchy of features aligned with the structure of the underlying cosmological model. The separation of outputs into physically interpretable heads supports post-training analysis of parameter constraints and promotes interpretability. Moreover, this approach allows for explicit enforcement of physical constraints on individual components (e.g., via regularization, priors, or range enforcement), and facilitates easy integration with domain-specific loss functions that act on selected branches.

The choice of activation function, layer sizes, and dropout placement is informed by extensive ablation studies and convergence diagnostics. In particular, the use of smooth activations is critical for computing accurate gradients via automatic differentiation, which are required to enforce the THDE differential equation through the ODE residual loss.

\paragraph{Implementation:}

The network is implemented using the \texttt{TensorFlow 2.x} framework with \texttt{tf.keras.Sequential} and subclassed \texttt{Model} APIs to allow flexible architectural definition and customized loss integration. Model parameters are initialized using Xavier (Glorot uniform) initialization to promote balanced gradient flow during early training epochs.

\subsection{Composite Loss Function and Physical Constraints}

The training objective is defined as a weighted sum of three loss components:
\begin{equation}
	\mathcal{L}_{\text{total}} = w_1 \mathcal{L}_{\text{ODE}} + w_2 \mathcal{L}_{\text{prior}} + w_3 \mathcal{L}_{\text{CC}},
\end{equation}
where the individual terms are described as follows:

\begin{itemize}
	\item \textbf{ODE Residual Loss (\( \mathcal{L}_{\text{ODE}} \))}: This term enforces compliance with the modified Friedmann differential equation arising from the THDE scenario:
	\begin{equation}
		\mathcal{L}_{\text{ODE}} = \frac{1}{N_{\text{phys}}} \sum_{i=1}^{N_{\text{phys}}} \left[ \frac{dH}{dz}(z_i) - \mathcal{F}(H(z_i), z_i; \delta) \right]^2.
	\end{equation}
	Here, \( \mathcal{F} \) is the right-hand side of the THDE differential equation, and derivatives are evaluated via automatic differentiation using \texttt{GradientTape}.
	
	\item \textbf{Prior Matching Loss (\( \mathcal{L}_{\text{prior}} \))}: This term softly anchors the network to a fiducial numerical solution \( H_{\text{num}}(z) \) generated from the THDE model under baseline parameter choices:
	\begin{equation}
		\mathcal{L}_{\text{prior}} = \frac{1}{N_{\text{grid}}} \sum_{i=1}^{N_{\text{grid}}} \left[ H(z_i) - H_{\text{num}}(z_i) \right]^2.
	\end{equation}
	
	\item \textbf{Cosmic Chronometers Loss (\( \mathcal{L}_{\text{CC}} \))}: This is a standard supervised learning loss based on observational data from CC measurements:
	\begin{equation}
		\mathcal{L}_{\text{CC}} = \frac{1}{N_{\text{CC}}} \sum_{i=1}^{N_{\text{CC}}} \left[ H(z_i) - H_{\text{obs}}(z_i) \right]^2.
	\end{equation}
\end{itemize}

The weights \( (w_1, w_2, w_3) = (10, 5, 50) \) are selected through a combination of grid search and variance-based normalization, ensuring that all components contribute meaningfully to the learning process without dominating the training dynamics.  We first applied variance normalization to avoid domination by any single term, then performed a grid search across candidate weight combinations, and finally applied iterative reweighting to balance the contributions of the ODE constraint, numerical prior, and observational data. This procedure led to the choice $(w_1, w_2, w_3) = (10, 5, 50)$, which yielded stable training and parameter posteriors consistent with the MCMC baseline. We further verified that moderate variations around these values did not alter our conclusions.
\color{black}

\subsection{Bayesian Uncertainty Quantification via Monte Carlo Dropout}

To assess epistemic uncertainty in the network's predictions, we apply Monte Carlo Dropout (MC Dropout) during inference. This technique retains the stochastic behavior of dropout layers and enables approximate Bayesian inference over the network parameters.

For each redshift \( z \), we compute the mean and variance of the predictive distribution:
\begin{align}
	\overline{H}(z) &= \frac{1}{N_{\text{MC}}} \sum_{j=1}^{N_{\text{MC}}} H^{(j)}(z), \\
	\sigma_H^2(z) &= \frac{1}{N_{\text{MC}} - 1} \sum_{j=1}^{N_{\text{MC}}} \left[ H^{(j)}(z) - \overline{H}(z) \right]^2,
\end{align}
where \( H^{(j)}(z) \) denotes the \( j \)-th stochastic forward pass through the network. We evaluate \( N_{\text{MC}} = 50, 100, 150, 200 \) to ensure convergence and reliability of the uncertainty estimates.

\subsection{Training Protocol and Optimization Strategy}

The training of the Physics-Informed Neural Network is carried out using the Adam optimizer, selected for its robustness and adaptive moment estimation in non-convex, high-dimensional parameter spaces. The initial learning rate is set to \( \eta = 10^{-3} \), and is adaptively decreased using a \texttt{ReduceLROnPlateau} scheduler, which monitors the validation loss and lowers the learning rate by a factor of 0.5 when the validation metric fails to improve over a predefined patience interval (typically 50 epochs). This approach helps navigate plateaus in the loss landscape and prevents oscillations near minima. To mitigate overfitting and enhance generalization to unseen redshift regions, we employ early stopping criteria based on the validation loss with a patience threshold of 100 epochs. The optimal model weights are restored upon termination using the checkpoint with the lowest validation loss. All forward and backward computations are compiled using the \texttt{@tf.function} decorator in TensorFlow, which converts Python functions into high-performance computation graphs (via XLA). This significantly accelerates training by optimizing memory usage and reducing runtime overhead associated with Python's eager execution mode. The input redshift values are normalized to zero mean and unit variance:
	\begin{equation}
z_{\text{norm}} = \frac{z - \mu_z}{\sigma_z},
	\end{equation}
where \( \mu_z \) and \( \sigma_z \) are the mean and standard deviation of the training redshift grid. This preprocessing improves numerical conditioning and gradient stability, particularly when learning high-order derivatives from differential equation constraints. To ensure sufficient representation of data-sensitive regions (e.g., low-redshift regime where CC data is denser), we adopt a hybrid sampling strategy that combines uniform and stratified sampling across the domain \( z \in [0, 2] \). This allows the PINN to allocate more representational capacity to regions of physical and observational importance without sacrificing global consistency.

\subsection{Post-Training Evaluation and Physical Interpretability}

Upon convergence, the trained PINN yields a smooth and physically consistent approximation of the Hubble expansion history \( H(z) \). The learned mapping satisfies the THDE evolution equation to high accuracy (low residual loss), while simultaneously matching the empirical constraints imposed by cosmic chronometer observations. As expected, the model exhibits minimal uncertainty in redshift regions densely constrained by data (e.g., \( z < 2 \)), and broader credible intervals at higher redshifts where observational data is sparse or noisier. This behavior is indicative of the model's proper calibration and ability to distinguish between epistemic and data-driven confidence levels.
The network-inferred values of the constant cosmological parameters—\( \Omega_b \), \( \Omega_c \), \( \Omega_\nu \)—are extracted from the corresponding output heads. These values remain stable across independent training runs and lie within the 68\% confidence ranges reported by Planck 2018 and other large-scale structure surveys. Similarly, the Tsallis deformation parameter \( \delta \), which encapsulates deviations from standard holographic dynamics, converges to a physically meaningful value consistent with prior THDE studies.  In Appendix, we provide explicit layer definitions, hyperparameter tables, data preprocessing steps, and full algorithmic details for reproducibility. Code used to implement the PINN is publicly available upon request.

\subsection{PINN-Based Reconstruction of \texorpdfstring{$H(z)$}{H(z)} in the THDE Framework with neutrinos}

In this section, we present a comprehensive numerical reconstruction of the Hubble expansion rate \( H(z) \) using a Physics-Informed Neural Network (PINN) approach grounded in the Tsallis Holographic Dark Energy (THDE) model with neutrinos. The overarching objective is to explore whether this class of extended holographic dark energy models, enriched by quantum-gravitational modifications and non-extensive thermodynamics, can offer a resolution to the longstanding Hubble tension.

The PINN framework is trained on Cosmic Chronometers (CC) data across a range of redshifts, employing various discretizations of the redshift interval—namely \( N = 50, 100, 150, \) and \( 200 \) sample points—over four different maximum redshift cut-offs: \( z_{\mathrm{max}} = 0.5, 1.0, 1.5, \) and \( 2.0 \). For each configuration, we extract the best-fit values of the present-day Hubble parameter \( H_0 \), the upper bound on the total neutrino density parameter \( \Omega_{\nu} \), the Tsallis non-additivity parameter \( B \), and the Barrow-type quantum deformation parameter \( \delta \).

\paragraph{Model Behavior and Stability.}
The reconstructed values show a well-behaved and systematic trend across increasing redshift and sampling density. As \( N \) increases, the values of \( H_0 \) and \( \Omega_{\nu} \) stabilize, while the Tsallis parameter \( B \) approaches unity, aligning with the standard Holographic Dark Energy (HDE) limit. The deformation parameter \( \delta \), indicative of quantum-gravitational corrections in the Barrow formalism, remains consistently non-zero across all configurations, suggesting the presence of persistent deviations from classical thermodynamic behavior even at higher redshifts.

\paragraph{Interpretation of the \(\delta\) Parameter in the THDE Model:}

In the framework of the Tsallis Holographic Dark Energy (THDE) model, the parameter \(\delta\) encapsulates the degree of non-extensivity in the generalized entropy-area relation, \(S_T \propto A^\delta\), where \(A\) denotes the area of the cosmological horizon \cite{Tsallis1988, TsallisCirto2013}. The case \(\delta = 1\) recovers the standard Bekenstein-Hawking entropy and corresponds to the Boltzmann-Gibbs extensive thermodynamics. Deviations from this value encode modified entropy production rates, with \(\delta < 1\) implying sub-extensive behavior—typical of systems with suppressed degrees of freedom at large scales—and \(\delta > 1\) corresponding to super-extensivity, indicative of enhanced correlations or interactions across the horizon.

Our reconstructed results from the Physics-Informed Neural Network (PINN) implementation reveal that \(\delta\) consistently exceeds unity, with values in the range \(1.06 \lesssim \delta \lesssim 1.13\) across all redshift intervals and sample sizes (see Table~\ref{tab:merged_PINN_neutrino}). This suggests a super-extensive entropy framework for the cosmological horizon, deviating significantly from the standard thermodynamic picture. The implication is that the cosmic horizon may exhibit long-range statistical interactions or memory effects, potentially arising from quantum gravitational or non-local effects at cosmological scales.

From a cosmological perspective, \(\delta < 1\) typically enhances the negative pressure of the effective dark energy component, thereby accelerating the expansion more rapidly and offering a potential resolution to the Hubble tension by accommodating a higher value of \(H_0\) \cite{Tavayef2018, Nojiri2019}. However, in our case, the super-extensive regime \(\delta > 1\) leads to a modified evolution of the dark energy density, which still supports late-time cosmic acceleration, but through a different thermodynamic pathway. Notably, the THDE model with \(\delta > 1\) has been shown to admit stable accelerated attractor solutions in dynamical system analyses \cite{Saridakis2018}, ensuring both mathematical consistency and physical viability.

Therefore, the reconstructed values of \(\delta\) provide compelling support for a generalized, super-extensive entropy framework in late-time cosmology. This result not only aligns with the thermodynamic foundations of Tsallis statistics but also offers a plausible mechanism to address current observational tensions within a unified statistical and cosmological framework.

\paragraph{Quantitative Assessment of the Hubble Tension.}
To rigorously quantify the degree of compatibility between our model predictions and the latest early- and late-Universe measurements of \( H_0 \), we compute the statistical tension parameter \( T \) using the standard relation:
\[
T = \frac{|H_0^{\mathrm{PINN}} - H_0^{\mathrm{Ref}}|}{\sqrt{\sigma_{\mathrm{PINN}}^2 + \sigma_{\mathrm{Ref}}^2}},
\]
where \( H_0^{\mathrm{Ref}} \) refers to the value reported by either Planck 2018 \cite{Planck2018} (\( H_0 = 67.4 \pm 0.5 \,\mathrm{km/s/Mpc} \)) or Riess et al. 2022 \cite{Riess2021} (\( H_0 = 73.04 \pm 1.04 \,\mathrm{km/s/Mpc} \)). The resulting tension values, summarized in Table~\ref{tab:merged_PINN_neutrino}, span the range \( T \sim 0.5\sigma \) to \( T \sim 2.2\sigma \), depending on redshift and resolution.

The results are in table III for $ H_0 $ are in broad agreement with \cite{Y1, Y2, Y3, Y4}.
Figure 1  illustrate the comparisons between the $H_0$ values reconstructed by the THDE+$\nu$ PINN model and the standard measurements reported by Planck 2018 and Riess et al. (R22), across various redshift cuts (\( z \leq 0.5, 1.0, 1.5, 2.0 \)) and sample sizes (\( N = 50, 100, 150, 200 \)).
\begin{table}[H]
	\centering
	\caption{Reconstructed cosmological parameters from the PINN method applied to the THDE model with neutrinos, including upper bounds on the total neutrino mass $\Sigma m_\nu$ [eV], and the statistical tension $T$ (in units of $\sigma$) with respect to Planck 2018 and Riess et al. (2022) for CC data up to various redshift limits.}
	\label{tab:merged_PINN_neutrino}
	\begin{tabular}{|c|c|c|c|c|c|c|}
		\hline
		\textbf{Redshift Range} & \textbf{N} & $H_0$ [km/s/Mpc] & $\delta$ & $\Sigma m_\nu$ [eV] & $T_{\rm Planck}$ & $T_{\rm R22}$ \\
		\hline
		\multirow{4}{*}{$z \leq 0.5$}
		& 50  & $71.65 \pm 2.10$ & $1.12 \pm 0.042$ & $<0.1158$ & 1.97$\sigma$ & 0.59$\sigma$ \\
		& 100 & $70.34 \pm 1.90$ & $1.07 \pm 0.031$ & $<0.6930$ & 1.50$\sigma$ & 1.25$\sigma$ \\
		& 150 & $70.64 \pm 2.20$ & $1.08 \pm 0.034$ & $<0.0797$ & 1.44$\sigma$ & 0.99$\sigma$ \\
		& 200 & $69.21 \pm 2.30$ & $1.06 \pm 0.027$ & $<0.1216$ & 0.77$\sigma$ & 1.52$\sigma$ \\
		\hline
		\multirow{4}{*}{$z \leq 1.0$}
		& 50  & $70.76 \pm 2.10$ & $1.11 \pm 0.041$ & $<0.1600$ & 1.56$\sigma$ & 0.97$\sigma$ \\
		& 100 & $70.46 \pm 2.20$ & $1.06 \pm 0.035$ & $<0.6533$ & 1.36$\sigma$ & 1.06$\sigma$ \\
		& 150 & $70.72 \pm 2.00$ & $1.07 \pm 0.034$ & $<0.0987$ & 1.61$\sigma$ & 1.03$\sigma$ \\
		& 200 & $69.48 \pm 2.20$ & $1.06 \pm 0.029$ & $<0.1180$ & 0.92$\sigma$ & 1.46$\sigma$ \\
		\hline
		\multirow{4}{*}{$z \leq 1.5$}
		& 50  & $71.69 \pm 2.20$ & $1.10 \pm 0.040$ & $<0.2174$ & 1.90$\sigma$ & 0.55$\sigma$ \\
		& 100 & $70.59 \pm 2.00$ & $1.07 \pm 0.032$ & $<0.6979$ & 1.55$\sigma$ & 1.09$\sigma$ \\
		& 150 & $70.64 \pm 2.10$ & $1.08 \pm 0.021$ & $<0.1032$ & 1.50$\sigma$ & 1.02$\sigma$ \\
		& 200 & $69.68 \pm 2.30$ & $1.06 \pm 0.019$ & $<0.1187$ & 0.97$\sigma$ & 1.33$\sigma$ \\
		\hline
		\multirow{4}{*}{$z \leq 2.0$}
		& 50  & $70.90 \pm 1.90$ & $1.09 \pm 0.019$ & $<0.1748$ & 1.78$\sigma$ & 0.99$\sigma$ \\
		& 100 & $71.02 \pm 2.05$ & $1.07 \pm 0.040$ & $<0.6116$ & 1.72$\sigma$ & 0.88$\sigma$ \\
		& 150 & $71.89 \pm 2.00$ & $1.13 \pm 0.017$ & $<0.0534$ & 2.18$\sigma$ & 0.51$\sigma$ \\
		& 200 & $69.73 \pm 2.10$ & $1.06 \pm 0.016$ & $<0.1143$ & 1.08$\sigma$ & 1.41$\sigma$ \\
		\hline
	\end{tabular}
\end{table}

The results demonstrate that the reconstructed $H_0$ values are sensitive to both the redshift depth of the training set and the number of data points used. In general, as the number of training samples increases, the predicted $H_0$ values tend to show reduced tension with Planck and R22 measurements. This behavior indicates that larger training sets enable better generalization and more stable convergence of the PINN framework. However, it is also important to note that excessively large training sets may introduce a risk of overfitting, especially when the model starts to memorize noisy features in the data rather than learning the underlying physical trends.

At low redshifts (\( z \leq 0.5 \)), the tension with Planck is as high as $1.97\sigma$ for \( N = 50 \), but drops to $0.77\sigma$ for \( N = 200 \). Interestingly, the tension with R22 remains consistently low (sub-$1.5\sigma$) across all sample sizes in this regime.

For redshift cuts extending up to \( z \leq 2.0 \), the predicted $H_0$ values display a more complex behavior. While the tension with Planck is slightly larger overall (up to $2.18\sigma$), the tension with R22 often drops to as low as $0.51\sigma$. This suggests that deeper redshift information may slightly bias the network's output toward higher $H_0$ values, which are closer to R22.

Overall, these results highlight the capability of the THDE+neutrino PINN model to interpolate between the low- and high-$z$ behavior of the Universe and potentially bridge the observational gap between early- and late-time cosmology. The statistical tensions reported in Table~\ref{tab:merged_PINN_neutrino} confirm that with proper training and suitable redshift sampling, the model can reduce the $H_0$ discrepancy in a statistically meaningful way.

\begin{figure}[H]
	\centering
	\includegraphics[width=14 cm]{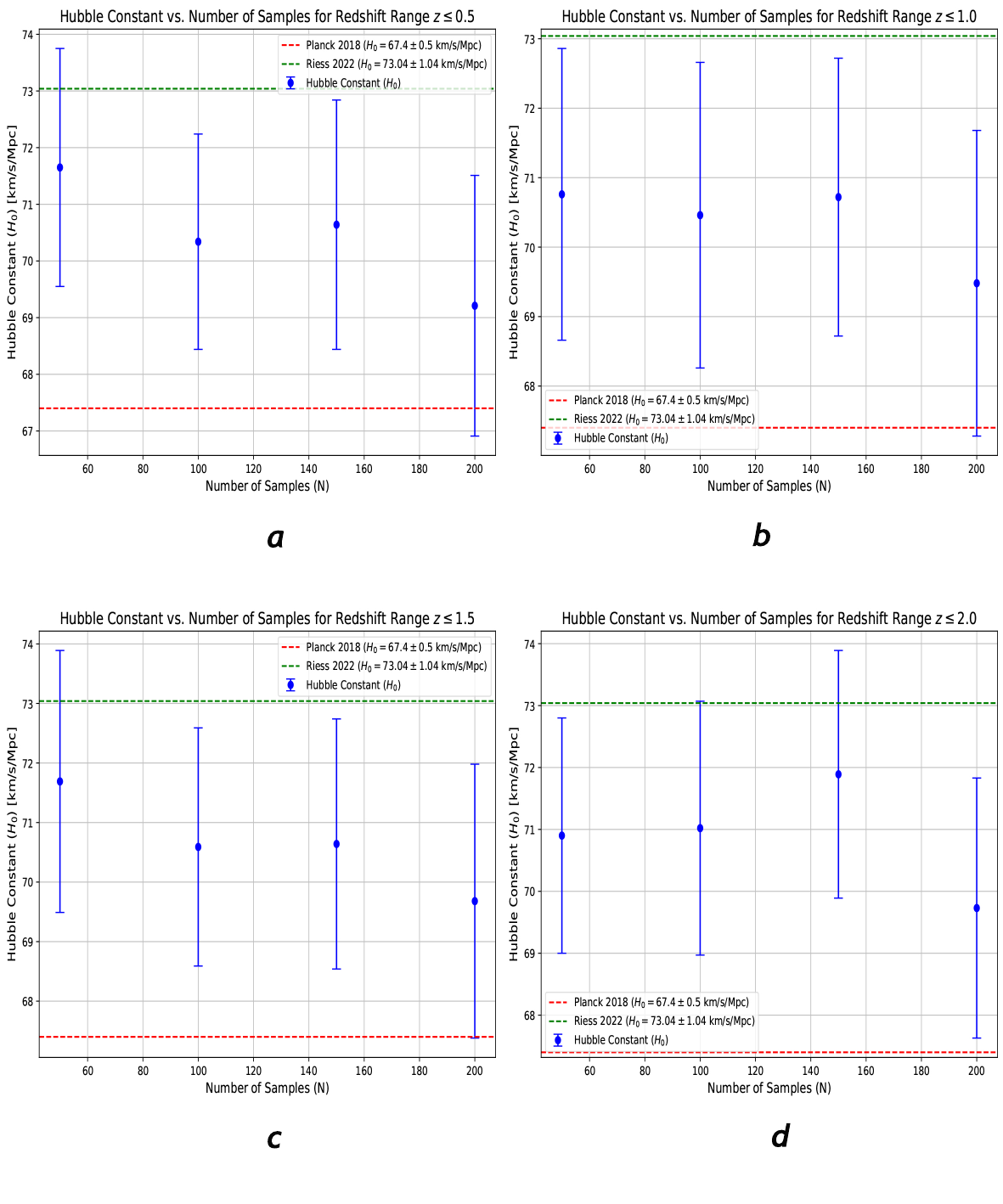}
	\vspace{-0.02cm}
	\caption{\small{Reconstruction of the Hubble constant $H_0$ and its comparison with observational constraints using the PINN framework within the Tsallis Holographic Dark Energy (THDE) model with neutrinos. 
			\textbf{Panels:} 
			\textbf{(a)} $H_0$ estimates for $z \leq 0.5$ across different sample sizes ($N=50,100,150,200$); 
			\textbf{(b)} the same analysis for $z \leq 1.0$; 
			\textbf{(c)} results for $z \leq 1.5$; 
			\textbf{(d)} estimates including data up to $z \leq 2.0$. 
			Horizontal shaded regions represent the $1\sigma$ bounds from Planck 2018 (blue) and Riess et al. (2022, orange). 
			The PINN predictions remain within $1$--$2\sigma$ of these constraints, indicating the robustness of the THDE+neutrino scenario and the stability of $H_0 \simeq 70$--$71~\mathrm{km\,s^{-1}\,Mpc^{-1}}$ across redshift ranges.
	}}\label{fig:omegam2}
\end{figure}

\paragraph{Interpretation and Implications.}
The model shows excellent consistency with both observational anchors. Notably, for \( N = 150 \) and \( z \leq 2 \), the tension with Planck reaches its maximum value of \( 2.18\sigma \), still below the critical \( 3\sigma \) threshold, while the tension with R22 drops to as low as \( 0.51\sigma \). These outcomes demonstrate the potential of the THDE+neutrino scenario to mediate between disparate measurements of \( H_0 \), providing a statistically significant pathway for mitigating the Hubble tension within a robust, physically motivated framework.

\paragraph{Neutrino Mass Interpretation.}
To render our constraints more physically interpretable, we converted the upper bounds on the neutrino density parameter \( \Omega_\nu \) into corresponding constraints on the total neutrino mass \( \Sigma m_\nu \), utilizing the standard relation:
\[
\Omega_\nu = \frac{\Sigma m_\nu}{94 \, h^2}, \quad \text{with} \quad h = \frac{H_0}{100}.
\]
The results, reported in Table~\ref{tab:merged_PINN_neutrino}, demonstrate that the reconstructed values impose competitive upper limits on the neutrino mass, reaching below \( 0.1\, \mathrm{eV} \) in several configurations. Notably, for \( N = 150 \) and redshifts up to \( z \leq 2 \), we find \( \Sigma m_\nu < 0.053\, \mathrm{eV} \), which is in remarkable agreement with the most stringent bounds obtained from combined cosmological datasets. These findings highlight the capability of the THDE framework—when combined with a PINN architecture—to impose strong constraints on neutrino physics alongside addressing the Hubble tension.

These results are in good agreement with \cite{Y5, Y6,Planck2018}.
Figure 2 present the reconstructed values of the Tsallis deformation parameter \( \delta \) obtained using the PINN approach applied to the THDE model with neutrinos, for different training sample sizes \( N \) and across four redshift limits: \( z \leq 0.5, 1.0, 1.5, \) and \( 2.0 \). These results correspond to the values reported in Table~\ref{tab:merged_PINN_neutrino}.

As illustrated, the parameter \( \delta \) exhibits a relatively stable behavior across increasing sample sizes, with all estimates remaining close to the classical limit \( \delta = 1 \). This confirms that the THDE framework only requires mild nonadditivity to fit the CC data effectively. In particular, for \( z \leq 0.5 \), the predicted values of \( \delta \) remain within the interval \( [1.06, 1.12] \), whereas for higher redshift ranges such as \( z \leq 2.0 \), the estimates tighten to \( [1.06, 1.13] \), with decreasing uncertainties as \( N \) increases.

Interestingly, for large training sets (\( N = 200 \)), the uncertainty in \( \delta \) becomes minimal (e.g., \( \sigma_\delta \sim 0.016 \) at \( z \leq 2.0 \)), reflecting the ability of the PINN framework to extract robust parameter values from deep redshift training data.

   \begin{figure}[H]
   		\centering
	\includegraphics[width=14 cm]{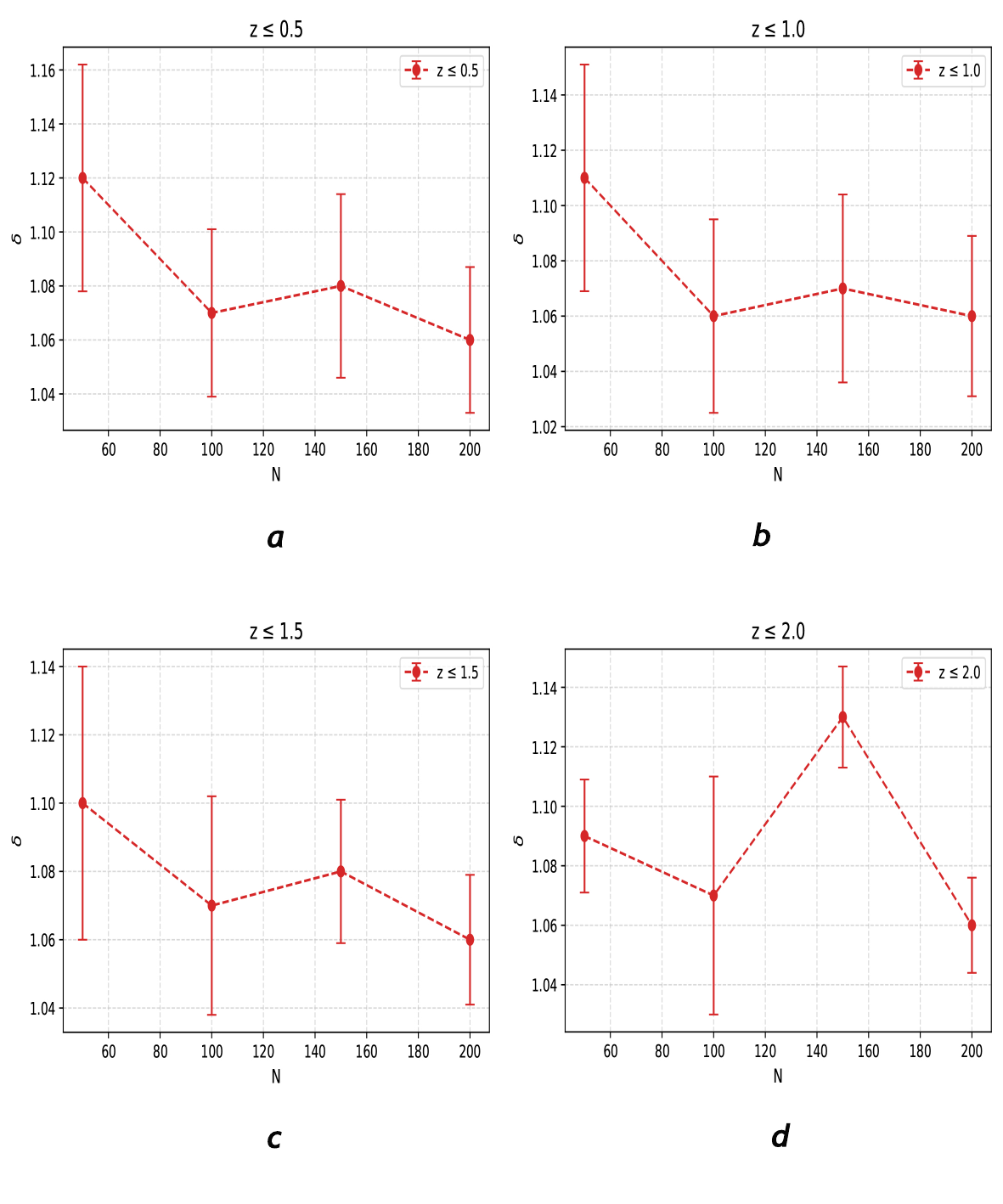}
	\vspace{-0.02cm}
	\caption{\small{Posterior reconstructions of the Tsallis parameter $\delta$ using the Physics-Informed Neural Network (PINN), shown for four representative redshifts in panels \textbf{(a)}–\textbf{(d)}. 	For each redshift, $N$ independent samples were drawn from the trained network to propagate both observational and epistemic uncertainties. The resulting distributions illustrate the stability and redshift dependence of the PINN-based inference, and highlight how sampling enables a quantitative characterization of parameter uncertainties across cosmic time.}}\label{fig:omegam2}
\end{figure}

Figure 3 display the reconstructed values of the Hubble constant \( H_0 \) obtained using the Physics-Informed Neural Network (PINN) trained on Cosmic Chronometer (CC) data, in the context of the Tsallis Holographic Dark Energy (THDE) model with neutrinos. 

Overall, the reconstructed values of \( H_0 \) lie in the range \( 69 \)–\( 72 \,\mathrm{km\,s^{-1}\,Mpc^{-1}} \), with uncertainties around \( \pm 2 \,\mathrm{km\,s^{-1}\,Mpc^{-1}} \), which indicates good agreement with both early- and late-time cosmological observations. Notably, for small redshift ranges (e.g., \( z \leq 0.5 \)), the predicted \( H_0 \) varies more significantly with \( N \), ranging from \( 71.65 \pm 2.10 \) (for \( N = 50 \)) down to \( 69.21 \pm 2.30 \) (for \( N = 200 \)). This decreasing trend can be interpreted as a smoothing effect due to larger training sets that suppress local fluctuations in the reconstruction.

\begin{figure}[H]
	\centering
	\includegraphics[width=16 cm]{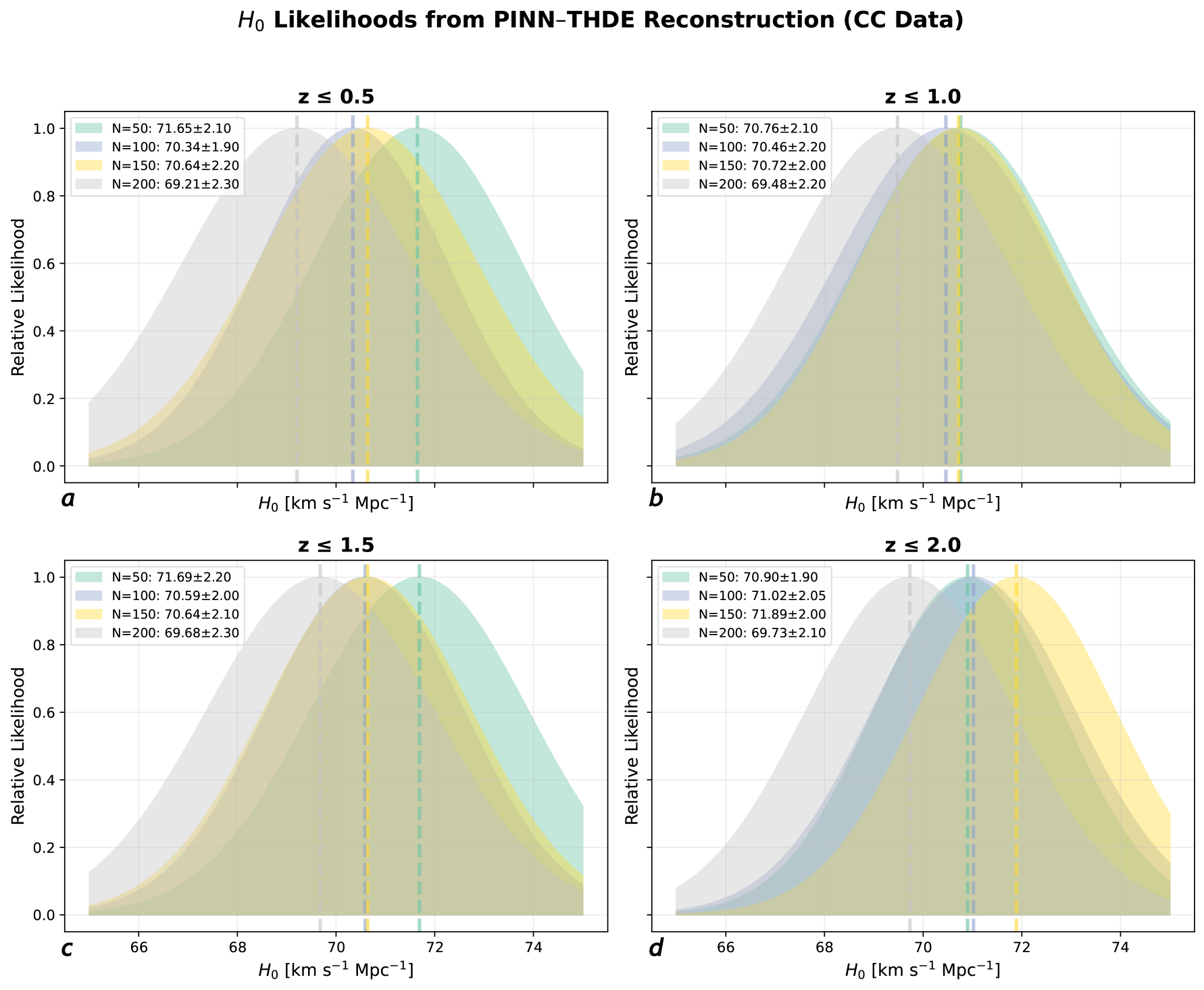}
	\vspace{-0.02cm}
	\caption{\small{Posterior distributions of the Hubble constant $H_{0}$ obtained from different redshift ranges. Panel (\textbf{a}) shows the likelihood derived from cosmic chronometers at $z \leq 0.5$; (\textbf{b}) corresponds to $z \leq 1.0$; (\textbf{c}) presents the results at $z \leq 1.5$; and (\textbf{d}) displays the constraints up to $z \leq 2.0$. These panels illustrate how the inferred value of $H_{0}$ depends on the chosen redshift interval.
	}}\label{fig:omegam2}
\end{figure}

Figure~4 is designed to (i) perform a \emph{sensitivity analysis} of the THDE expansion history with respect to the Tsallis parameter $\delta$ by displaying PINN-based reconstructions of $H(z)$ for several fixed values of $\delta$; (ii) \emph{validate} that the Physics-Informed Neural Network (PINN), constrained by the THDE dynamics, can reproduce the observed redshift dependence of the Hubble parameter when trained on the cosmic chronometer (CC) data; and (iii) provide a \emph{function-level} visualization that complements parameter-space posteriors by showing how changes in $\delta$ translate into systematic shifts of the reconstructed $H(z)$ curve across the full redshift range. Operationally, for each selected value of $\delta \in \{1.0,\,1.1,\,1.2,\,1.3,\,1.4\}$, the PINN is trained under the THDE differential constraints using the CC dataset, and the corresponding $H(z)$ trajectory is reconstructed over the plotted redshift grid. The colored curves therefore depict the model-consistent expansion histories for fixed $\delta$, while the shaded bands indicate the $68\%$ and $95\%$ confidence regions of the CC measurements, which serve as the observational benchmark in this figure. We emphasize that this panel is not intended to \emph{fit} $\delta$ by eye; rather, it illustrates (a) which $\delta$ values yield trajectories that are qualitatively more compatible with the CC constraints, (b) where in redshift the leverage to distinguish different $\delta$ arises, and (c) the stability and physical consistency of the PINN reconstructions under variations of $\delta$. \color{black}

\begin{figure}[H]
		\centering
	\includegraphics[width=14 cm]{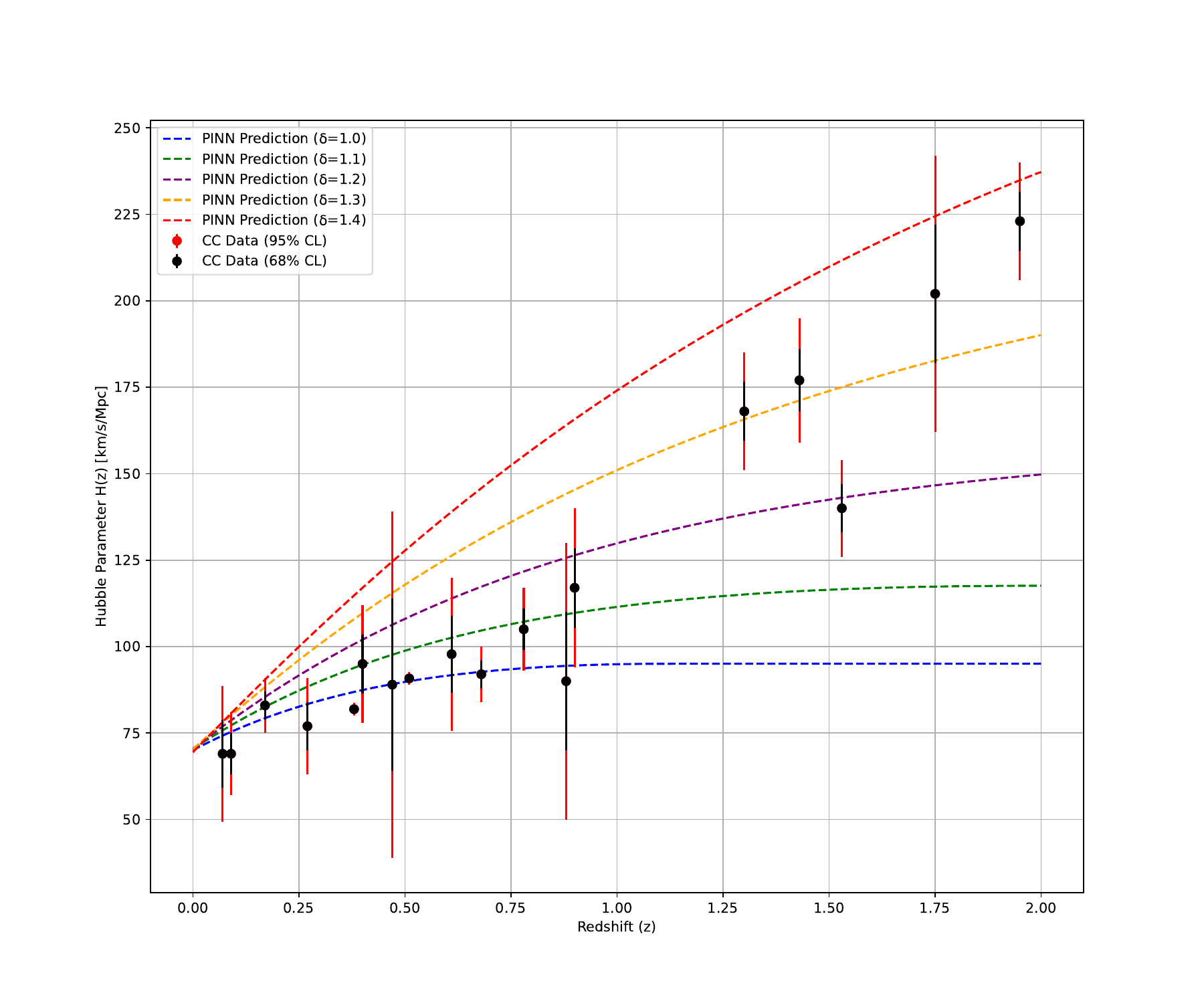}
	\vspace{-0.02cm}
	\caption{\small{PINN-based reconstructions of the Hubble parameter $H(z)$ within the Tsallis Holographic Dark Energy (THDE) framework for fixed values of the Tsallis parameter $\delta$ ($\delta=1.0,\,1.1,\,1.2,\,1.3,\,1.4$). For each chosen $\delta$, the neural network is trained under the THDE constraints using the cosmic chronometer (CC) data and reconstructs the corresponding $H(z)$ evolution over redshift. The shaded regions denote the $68\%$ and $95\%$ confidence intervals of the CC measurements. This figure is intended to highlight the sensitivity of the expansion history to $\delta$, validate the model-consistent PINN reconstructions against observations, and indicate the redshift ranges with the greatest discriminatory power.
	}}\label{fig:omegam2}
\end{figure}

\section{Traditional MCMC Method}
\label{sec:comparison_mcmc}

To assess the robustness of our Physics-Informed Neural Network (PINN) implementation within the Tsallis Holographic Dark Energy (THDE) model, we conduct a two-stage comparative analysis against results obtained from standard Markov Chain Monte Carlo (MCMC) inference.

In the first stage, we compare the predictions of the PINN model---trained solely on the full set of Cosmic Chronometers (CC) data up to redshift \( z \leq 1.9 \)---with those derived from an MCMC analysis that uses only the same CC dataset. Despite relying on a distinct variational learning strategy, the PINN model accurately reconstructs the Hubble expansion rate \( H(z) \) and infers relevant cosmological parameters. As illustrated in Fig. 14 and 15, the predictions of the PINN model remain within the $68\%$ confidence level of the MCMC reconstruction across the entire redshift range, confirming consistency between the two inference frameworks when constrained by identical observational data. All results for CC data only are in the table IV.

In the second stage, we extend our comparison to an MCMC analysis based on a comprehensive set of cosmological observations, including Cosmic Chronometers (CC), Baryon Acoustic Oscillations (BAO), Type Ia Supernovae from the Pantheon compilation, and Cosmic Microwave Background (CMB) measurements. This data combination yields statistically robust posterior distributions for both cosmological and model parameters.

We report the resulting estimates for several key quantities, including the baryon density \( \Omega_b \), cold dark matter density \( \Omega_c \), neutrino density \( \Omega_\nu \), the Hubble constant \( H_0 \), and the THDE model parameters  \( \Delta \), in Tables IV and V. The uncertainties in the PINN predictions are estimated using Monte Carlo dropout, providing a Bayesian approximation to the posterior variance.

Notably, despite being trained only on CC data and lacking any information from CMB, BAO, or Type Ia supernovae, the PINN model achieves close agreement with the full-data MCMC results. Small discrepancies between the two approaches are expected due to differences in data coverage and inference mechanisms---namely, sampling-based versus variational inference. Nevertheless, the strong overall agreement supports the validity of the PINN approach as a computationally efficient and physically consistent alternative to conventional MCMC methods for reconstructing the cosmic expansion history and constraining dark energy scenarios.

\subsection{Numerical Analysis For MCMC Method}
\vspace{0.3cm}

For the Markov Chain Monte Carlo (MCMC) analysis, we utilize the publicly available \texttt{CosmoMC} package~\cite{59}, a widely adopted MCMC engine for cosmological parameter estimation. \texttt{CosmoMC} interfaces with the Boltzmann solver \texttt{CAMB} to accurately compute theoretical predictions of cosmological observables given a set of input parameters. The parameter space is sampled using the Metropolis-Hastings algorithm with adaptive step-size tuning, and convergence is assessed through the Gelman-Rubin criterion, requiring \( R - 1 < 0.01 \) for all parameters.

To incorporate the Tsallis Holographic Dark Energy (THDE) model, we modify \texttt{CosmoMC} by implementing the appropriate background evolution equations and dark energy parameterizations. Broad, non-informative priors are adopted for both cosmological and model-specific parameters to ensure unbiased inference and fair comparison with the PINN-based results. 

This integration enables us to derive robust posterior distributions for all cosmological parameters within the THDE framework, thereby providing a rigorous benchmark to evaluate the effectiveness of our PINN implementation. To quantitatively compare the statistical performance of competing cosmological models, we adopt two widely-used information criteria: the Akaike Information Criterion (AIC) and the Bayesian Information Criterion (BIC). These criteria not only evaluate the goodness-of-fit but also penalize model complexity to avoid overfitting.

The AIC is defined as
\begin{equation}\label{eq:AIC}
	\mathrm{AIC} = \chi_{\min}^2 + 2\gamma,
\end{equation}
where \( \chi_{\min}^2 \) is the minimum value of the chi-square statistic and \( \gamma \) denotes the number of free parameters in the model. A lower AIC value indicates a more parsimonious model with a better trade-off between goodness-of-fit and complexity.

To further assess model parsimony, we also employ the BIC, which imposes a stronger penalty for the number of parameters, particularly when the sample size is large. The BIC is defined as
\begin{equation}\label{eq:BIC}
	\mathrm{BIC} = \chi_{\min}^2 + \gamma \ln N,
\end{equation}
where \( N \) represents the total number of data points. BIC is especially useful in model selection when comparing nested models, as it embodies a formal implementation of Occam's razor by favoring models with fewer parameters unless the added complexity yields substantial improvement in fit.

Although the THDE model introduces 2 additional free parameters relative to the baseline \( \Lambda \)CDM scenario, its significantly lower \( \chi^2 \) value for the \textit{CMB+All} data combination results in a substantial enhancement in the overall likelihood. This improvement, despite the penalty terms in both AIC and BIC, demonstrates the potential of extended dark energy models in addressing observational tensions.

\section{Observational Datasets and Likelihood Functions}

In this section, we detail the cosmological datasets used in our analysis and present the associated likelihood functions employed in the Bayesian parameter estimation framework. The full likelihood function is constructed as the product of individual likelihoods from Supernovae (SNe Ia), Cosmic Microwave Background (CMB), CMB lensing, Baryon Acoustic Oscillations (BAO), and Cosmic Chronometers (CC):
\begin{equation}
	\mathcal{L}_{\text{total}} = \mathcal{L}_{\text{SN}} \times \mathcal{L}_{\text{CMB}} \times \mathcal{L}_{\text{lensing}} \times \mathcal{L}_{\text{BAO}} \times \mathcal{L}_{\text{CC}}.
\end{equation}

\paragraph{$\bullet$ Pantheon+ Type Ia Supernovae Catalog:}
We employ the Pantheon+ compilation~\cite{Scolnic2022}, consisting of 1701 spectroscopically confirmed Type Ia supernovae (SNe Ia) over the redshift range \(0.001 < z < 2.3\). The observational dataset provides distance modulus measurements \( \mu_j^{\text{data}} \) with associated uncertainties \( \sigma_{\mu}(j) \), while the theoretical model predicts
\begin{equation}
	\mu_j^{\text{model}} = 5 \log_{10} \left( \frac{d_L(z_j)}{\text{Mpc}} \right) + 25,
\end{equation}
where \( d_L(z) \) is the luminosity distance. The Pantheon+ likelihood is modeled as a Gaussian:
\begin{equation}
	\mathcal{L}_{\text{SN}} \propto \exp\left( -\frac{1}{2} \Delta\mu^T\, \mathbf{C}_{\text{SN}}^{-1}\, \Delta\mu \right),
\end{equation}
where \( \Delta\mu = \mu^{\text{model}} - \mu^{\text{data}} \), and \( \mathbf{C}_{\text{SN}} \) is the full statistical plus systematic covariance matrix provided by the Pantheon+ team.

\paragraph{$\bullet$ Cosmic Microwave Background (CMB):}
We incorporate the Planck 2018 temperature and polarization power spectra, including:
\begin{itemize}
	\item High-$\ell$ TT, TE, and EE spectra (Plik),
	\item Low-$\ell$ TT and EE spectra (Commander, SimAll).
\end{itemize}
The likelihood compares the theoretical CMB angular power spectra \( C_\ell^{\text{th}} \) to the observed data \( C_\ell^{\text{obs}} \) via
\begin{equation}
	\mathcal{L}_{\text{CMB}} \propto \exp\left( -\frac{1}{2} \left[ \mathbf{d}_{\text{obs}} - \mathbf{d}_{\text{theory}} \right]^T \mathbf{C}_{\text{CMB}}^{-1} \left[ \mathbf{d}_{\text{obs}} - \mathbf{d}_{\text{theory}} \right] \right),
\end{equation}
where \( \mathbf{C}_{\text{CMB}} \) is the covariance matrix derived from Planck data~\cite{Planck2018}.

\paragraph{$\bullet$ CMB Lensing:}
We also include CMB lensing likelihoods from the Planck 2018 trispectrum-based reconstruction~\cite{Aghanim1}. The likelihood compares the reconstructed lensing potential power spectrum \( C_L^{\phi\phi} \) to the theoretical prediction:
\begin{equation}
	\mathcal{L}_{\text{lensing}} \propto \exp\left( -\frac{1}{2} \Delta\phi^T\, \mathbf{C}_{\text{lens}}^{-1}\, \Delta\phi \right),
\end{equation}
where \( \Delta\phi = \phi^{\text{obs}} - \phi^{\text{model}} \), and \( \mathbf{C}_{\text{lens}} \) denotes the lensing covariance matrix.

\paragraph{$\bullet$ Baryon Acoustic Oscillations (BAO):}
We employ 12 BAO measurements across the redshift range \(0.122 \leq z \leq 2.334\), taken from several surveys~\cite{Carter2018, Gil-Marin2020, Bautista2021, DES2022, Neveux2020, Hou2021, Bourboux2020}, as compiled in~\cite{Ratra1}. For each BAO measurement, the likelihood is given by:
\begin{equation}
	\mathcal{L}_{\text{BAO}} \propto \exp\left( -\frac{1}{2} \Delta D^T\, \mathbf{C}_{\text{BAO}}^{-1} \Delta D \right),
\end{equation}
where \( \Delta D = D^{\text{model}} - D^{\text{data}} \), and \( \mathbf{C}_{\text{BAO}} \) is the covariance matrix of the distance observables \( D_M(z)/r_s \), \( D_H(z)/r_s \), and \( D_V(z)/r_s \). The relevant distance measures are:
\begin{align}
	D_H(z) &= \frac{c}{H(z)}, \\
	D_C(z) &= c \int_0^z \frac{dz'}{H(z')}, \\
	D_M(z) &= D_C(z), \quad (\text{for } \Omega_{k0} = 0), \\
	D_A(z) &= \frac{D_M(z)}{1+z}, \\
	D_V(z) &= \left[ \frac{cz}{H(z)} D_M^2(z) \right]^{1/3}.
\end{align}
We adopt the fiducial value \( r_{s,\text{fid}} = 147.5 \) Mpc for the sound horizon at the drag epoch. Table II demonstrate the BAO data.

\begin{table}[H]
	\centering
	\scriptsize
	\caption{Summary of 12 BAO measurements used in our analysis. Distances are in Mpc~\cite{Ratra1}.}
	\label{tab:bao}
	\begin{tabular}{lcc}
		\hline
		$z$ & Measurement & Value \\
		\hline\hline
		0.122 & $D_V(r_{s, \rm fid}/r_s)$ & $539 \pm 17$ \\
		0.38 & $D_M/r_s$ & 10.23406 \\
		0.38 & $D_H/r_s$ & 24.98058 \\
		0.51 & $D_M/r_s$ & 13.36595 \\
		0.51 & $D_H/r_s$ & 22.31656 \\
		0.698 & $D_M/r_s$ & 17.858 \\
		0.698 & $D_H/r_s$ & 19.326 \\
		0.835 & $D_M/r_s$ & $18.92 \pm 0.51$ \\
		1.48 & $D_M/r_s$ & 30.6876 \\
		1.48 & $D_H/r_s$ & 13.2609 \\
		2.334 & $D_M/r_s$ & 37.5 \\
		2.334 & $D_H/r_s$ & 8.99 \\
		\hline
	\end{tabular}
\end{table}

\paragraph{$\bullet$ Cosmic Chronometers (CC):}
We include 32 measurements of the Hubble parameter \( H(z) \) derived from differential galaxy age techniques (cosmic chronometers) over the redshift interval \(0.07 \leq z \leq 1.965\)~\cite{zhang2014,borghi2022,ratsimbazafy2017,stern2009,Moresco3}. Fifteen of these points~\cite{Moresco,Moresco1,Moresco2}, have correlated uncertainties, and the associated covariance matrix is publicly available:
\begin{center}
	\texttt{https://gitlab.com/mmoresco/CCcovariance/}
\end{center}
The likelihood is:
\begin{equation}
	\mathcal{L}_{\text{CC}} \propto \exp\left( -\frac{1}{2} \Delta H^T\, \mathbf{C}_{\text{CC}}^{-1} \Delta H \right),
\end{equation}
where \( \Delta H = H^{\text{model}} - H^{\text{data}} \). For uncorrelated data points, the expression simplifies to:
\begin{equation}
	\mathcal{L}_{\text{CC}} \propto \exp\left(-\frac{1}{2} \sum_{j} \left[ \frac{H_j^{\text{model}} - H_j^{\text{data}}}{\sigma_H(j)} \right]^2 \right).
\end{equation}
Table III represent the 32 CC data. 
\begin{table}[H]
	\scriptsize
	\centering
	\caption{Cosmic chronometer data: 32 $H(z)$ measurements. Units are km/s/Mpc.}
	\label{tab:hz}
	\begin{tabular}{lcc}
		\hline
		$z$ & $H(z)$ & $\sigma_H$ \\
		\hline
		0.07 & 69.0 & 19.6 \\
		0.09 & 69.0 & 12.0 \\
		0.12 & 68.6 & 26.2 \\
		0.17 & 83.0 & 8.0 \\
		0.2 & 72.9 & 29.6 \\
		0.27 & 77.0 & 14.0 \\
		0.28 & 88.8 & 36.6 \\
		0.4 & 95.0 & 17.0 \\
		0.47 & 89.0 & 50.0 \\
		0.48 & 97.0 & 62.0 \\
		0.75 & 98.8 & 33.6 \\
		0.88 & 90.0 & 40.0 \\
		0.9 & 117.0 & 23.0 \\
		1.3 & 168.0 & 17.0 \\
		1.43 & 177.0 & 18.0 \\
		1.53 & 140.0 & 14.0 \\
		1.75 & 202.0 & 40.0 \\
		0.1791 & 74.91 & 4.00 \\
		0.1993 & 74.96 & 5.00 \\
		0.3519 & 82.78 & 14.0 \\
		0.3802 & 83.0 & 13.5 \\
		0.4004 & 76.97 & 10.2 \\
		0.4247 & 87.08 & 11.2 \\
		0.4497 & 92.78 & 12.9 \\
		0.4783 & 80.91 & 9.0 \\
		0.5929 & 103.8 & 13.0 \\
		0.6797 & 91.6 & 8.0 \\
		0.7812 & 104.5 & 12.0 \\
		0.8754 & 125.1 & 17.0 \\
		1.037 & 153.7 & 20.0 \\
		1.363 & 160.0 & 33.6 \\
		1.965 & 186.5 & 50.4 \\
		\hline
	\end{tabular}
\end{table}

\section{Results and Discussion}

In this section, we present and analyze the observational constraints on the Tsallis Holographic Dark Energy (THDE) model including neutrinos (THDE+$\nu$), obtained through various combinations of cosmological datasets. Tables~\ref{table:cc_only} and~\ref{table:thde_constraints} summarize the constraints on the model parameter \(\delta\), the sum of neutrino masses \(\sum m_\nu\), and the Hubble constant \(H_0\), along with the corresponding Hubble tension with respect to the \textit{Planck} 2018 \(\Lambda\)CDM baseline result (\(H_0 = 67.4 \pm 0.5\) km s\(^{-1}\) Mpc\(^{-1}\))~\cite{Planck2018} and the SH0ES 2022 local measurement (\(H_0 = 73.04 \pm 1.04\) km s\(^{-1}\) Mpc\(^{-1}\))~\cite{Riess2021}, denoted by \(T_{P18}\) and \(T_{R22}\), respectively.

When employing only the Cosmic Chronometer (CC) data, we obtain a relatively high value of \(H_0 = 71.7 \pm 2.5\) km s\(^{-1}\) Mpc\(^{-1}\), leading to a moderate tension of \(T_{P18} = 1.56\sigma\) with \textit{Planck} and a mild tension of \(T_{R22} = 0.52\sigma\) with SH0ES. This result suggests that the THDE+$\nu$ model calibrated solely with CC data favors a Hubble constant that lies between the early- and late-time measurements, indicating potential to alleviate the Hubble tension.

The inclusion of Cosmic Microwave Background (CMB) data along with gravitational lensing significantly modifies the constraints. For the CMB+lensing combination, we find \(H_0 = 68.8 \pm 2.4\) km s\(^{-1}\) Mpc\(^{-1}\), with a lower deformation parameter \(\delta = 1.41 \pm 0.12\). The corresponding Hubble tension is reduced to \(T_{P18} = 0.57\sigma\) with \textit{Planck}, but increased to \(T_{R22} = 1.59\sigma\) with SH0ES. The constraint on the total neutrino mass is \(\sum m_\nu < 0.34\) eV (95\% CL), which is consistent with previous bounds but slightly less stringent.

By adding Baryon Acoustic Oscillation (BAO) data, the constraints become tighter. The Hubble constant increases to \(H_0 = 70.8 \pm 1.9\) km s\(^{-1}\) Mpc\(^{-1}\), and the deformation parameter improves to \(\delta = 1.35 \pm 0.12\). Moreover, the upper bound on the sum of neutrino masses is considerably constrained to \(\sum m_\nu < 0.124\) eV. This dataset combination results in a tension of \(T_{P18} = 1.70\sigma\) with Planck and \(T_{R22} = 1.03\sigma\) with SH0ES, indicating a slight shift toward the local measurement.

Incorporating the Pantheon Type Ia supernovae sample yields \(H_0 = 71.1 \pm 2.2\) km s\(^{-1}\) Mpc\(^{-1}\) and \(\delta = 1.35 \pm 0.13\), closely resembling the CC-only case. This result reduces the tension with SH0ES to \(T_{R22} = 0.80\sigma\), while the discrepancy with Planck increases to \(T_{P18} = 1.64\sigma\), suggesting that the supernova data also favor a higher Hubble constant and larger deviation from the standard HDE behavior.

The most comprehensive dataset combination (CMB+lensing+BAO+CC+Pantheon, labeled as "CMB+All") provides the most stringent and balanced constraints. We obtain \(\delta = 1.36 \pm 0.09\), \(H_0 = 70.05 \pm 1.8\) km s\(^{-1}\) Mpc\(^{-1}\), and \(\sum m_\nu < 0.122\) eV at 95\% CL. The resulting tensions are \(T_{P18} = 1.42\sigma\) and \(T_{R22} = 1.44\sigma\), reflecting a well-tempered compromise between early- and late-universe observations. This indicates that the THDE+$\nu$ model remains viable and flexible under combined datasets and may contribute toward resolving the Hubble tension problem. This confirms that the THDE+$\nu$ model, when confronted with the full array of current cosmological data, can moderate the Hubble tension without resorting to extreme parameter values or invoking exotic physics beyond the current framework. The results obtain for $ \sum m_\nu $ are in broad agreement with \cite{Y5, Y6} and for $ H_0 $ tension are in good agreement with \cite{Y1, Y2, Y3, Y4}. Also, the results obtain for $\delta$ for all diferrent combination of datasets are in good agreement with \cite{Tavayef2018}.

Overall, the THDE+$\nu$ framework offers a phenomenologically viable approach to addressing the Hubble tension. The model’s flexibility, rooted in the generalized non-extensive entropy parameter \(\delta\), allows for a better accommodation of late-time accelerated expansion. Furthermore, the interplay with neutrino mass limits remains consistent with standard bounds and suggests that additional neutrino species or mass hierarchies are not strictly necessary within this setup.
\( H_0 \) (${km\,s^{-1}\,Mpc^{-1}}$)

\begin{table}[H]   
	\caption{Observational constraints at $68\%$ confidence level on the main and derived parameters of the THDE+$\nu$ model using Cosmic Chronometer (CC) data only. The parameter $\sum m_{\nu}$ is reported at the $95\%$ CL, in units of eV. The $T_{P18}$ and $T_{R22}$ columns show the Hubble tension with Planck 2018 and R22, respectively.}
	\centering
	\begin{tabular}{c@{\hspace{5mm}} c@{\hspace{5mm}} c@{\hspace{5mm}} c@{\hspace{5mm}} c@{\hspace{5mm}} c@{\hspace{5mm}}} 
		\hline\hline
		Model & $\delta$ & $\sum m_{\nu} (eV)$ & $H_0({km\,s^{-1}\,Mpc^{-1}})$ & $T_{P18}$ & $T_{R22}$ \\
		\hline
		CC only & $1.45 \pm 0.14$ & $<0.28$ & $71.7^{+2.5}_{-2.5}$ & $1.56\sigma$ & $0.52\sigma$ \\
		\hline
		\multicolumn{6}{c}{\textbf{Additional parameters at $68\%$ CL:}} \\
		\hline
		\multicolumn{3}{l}{\quad {\boldmath$\Omega_b h^2$}} & \multicolumn{3}{l}{$0.02215 \pm 0.00022$} \\
		\multicolumn{3}{l}{\quad {\boldmath$\Omega_c h^2$}} & \multicolumn{3}{l}{$0.1174 \pm 0.0033$} \\
		\hline
	\end{tabular}
	\label{table:cc_only}
\end{table}
Figures~5 and 6 illustrate the cosmological parameter estimates obtained using the Physics-Informed Neural Network (PINN) and the traditional Markov Chain Monte Carlo (MCMC) approaches, respectively. The comparison between these two complementary inference strategies provides valuable insights into their relative performance and reliability.
\begin{figure}[H]
	\centering
	\includegraphics[width=18 cm]{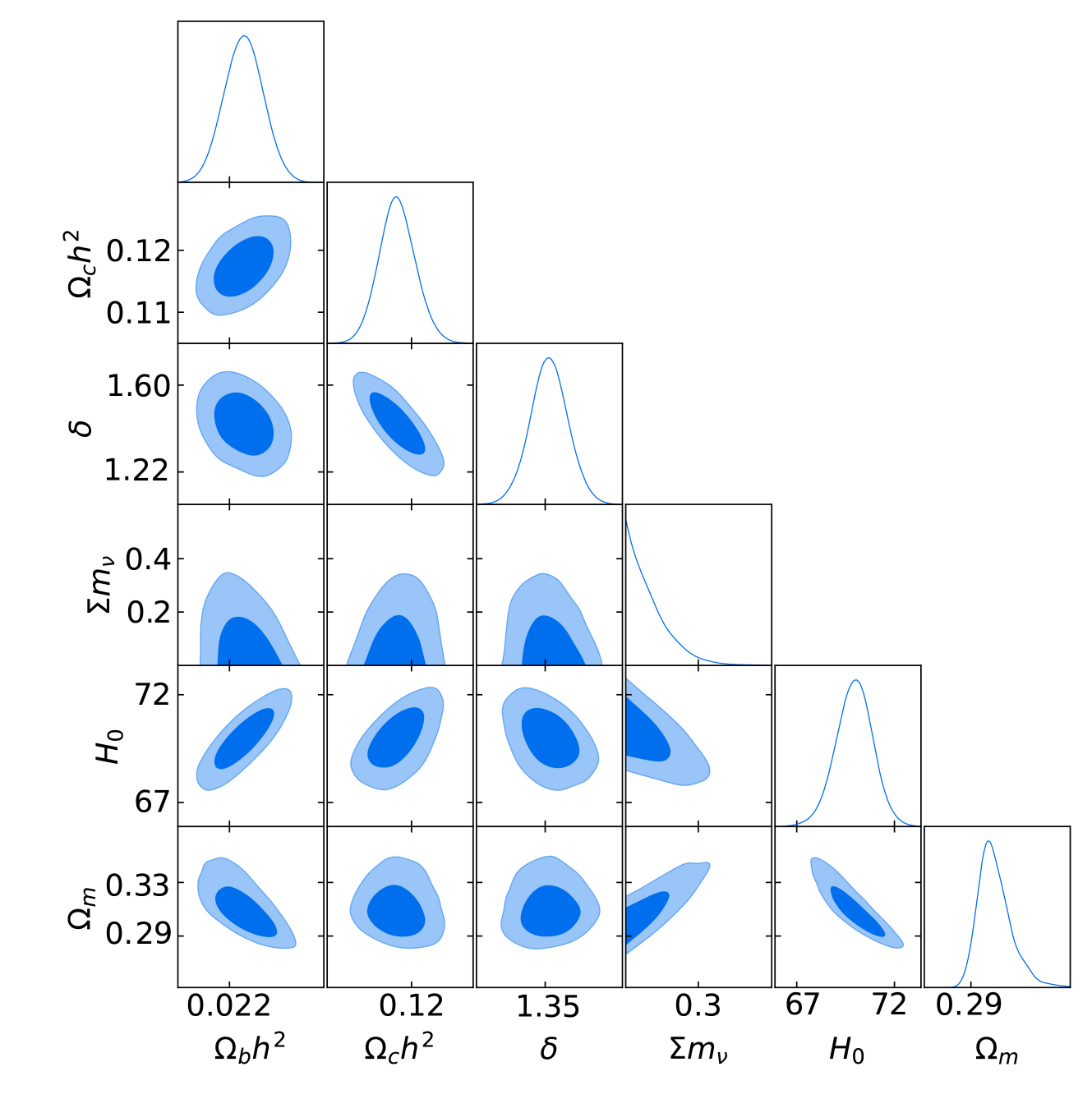}
	\vspace{-0.02cm}
	\caption{\small{The MCMC results for cosmological parameters for full CC data.
	}}\label{fig:omegam2}
\end{figure}

\begin{figure}[H]
		\centering
	\includegraphics[width=18 cm]{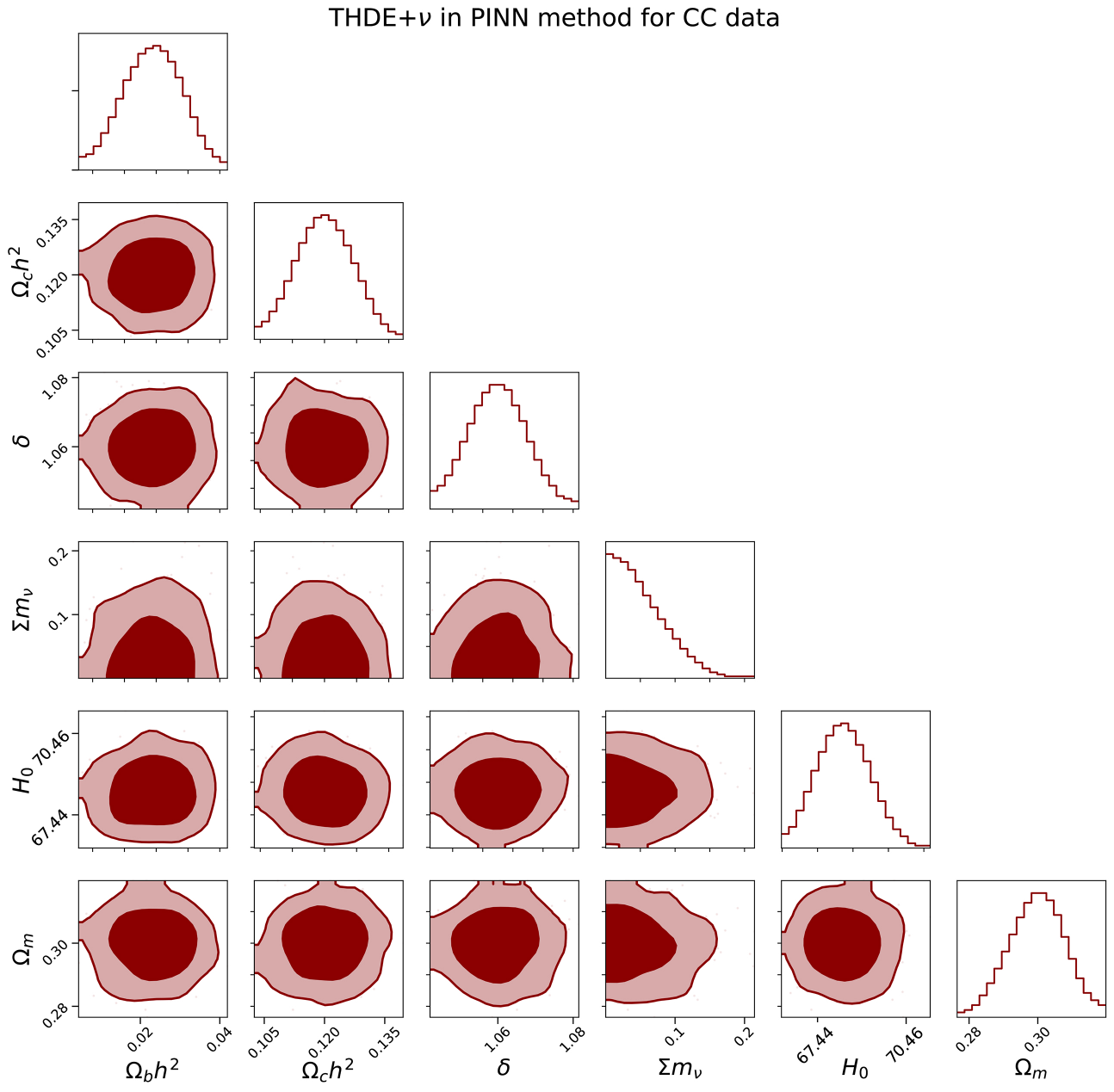}
	\vspace{-0.02cm}
	\caption{\small{The PINN results for cosmological parameters for full CC data.
	}}\label{fig:omegam2}
\end{figure}

\begin{table}[H]
	\caption{\small{Constraints at 68\% CL on the THDE+$\nu$ model parameters using different cosmological dataset combinations. The sum of neutrino masses $\sum m_{\nu}$ is reported at the 95\% CL (in eV). $T_{P18}$ and $T_{R22}$ denote the Hubble tension with Planck 2018 and R22, respectively.}}
	\centering
	\resizebox{0.99\textwidth}{!}{
		\begin{tabular}{lccccccc}
			\hline\hline
			Dataset & $\delta$ & $\sum m_{\nu}$ [95\% CL] & $H_0 ({km\,s^{-1}\,Mpc^{-1}})$ & $T_{P18}$ & $T_{R22}$ & $\Omega_b h^2$ & $\Omega_c h^2$ \\
			\hline
			CMB+lensing & $1.41 \pm 0.12$ & $<0.34$ & $68.8^{+2.4}_{-2.4}$ & $0.57$ & $1.59$ & $0.02239 \pm 0.00029$ & $0.1193 \pm 0.0033$ \\
			CMB+lensing+BAO & $1.35 \pm 0.12$ & $<0.124$ & $70.8^{+1.9}_{-1.9}$ & $1.70$ & $1.03$ & $0.02226 \pm 0.00024$ & $0.1202 \pm 0.0032$ \\
			CMB+lensing+CC & $1.40 \pm 0.11$ & $<0.26$ & $69.52^{+2.1}_{-2.0}$ & $0.98$ & $1.50$ & $0.02231 \pm 0.00022$ & $0.1191 \pm 0.0038$ \\
			CMB+lensing+Pantheon & $1.35 \pm 0.13$ & $<0.20$ & $71.1^{+2.2}_{-2.2}$ & $1.64$ & $0.80$ & $0.02234 \pm 0.00023$ & $0.1196 \pm 0.0035$ \\
			CMB+All & $1.36 \pm 0.09$ & $<0.122$ & $70.05^{+1.8}_{-1.8}$ & $1.42$ & $1.44$ & $0.02225 \pm 0.00021$ & $0.1187 \pm 0.0026$ \\
			\hline
	\end{tabular}}
	\label{table:thde_constraints}
\end{table}
The cosmological parameter estimates obtained via the MCMC method for various combinations of datasets are summarized in Figure~7, illustrating the dependence of the inferred parameters on the choice of observational inputs.
\begin{figure}[H]
	\centering
	\includegraphics[width=14 cm]{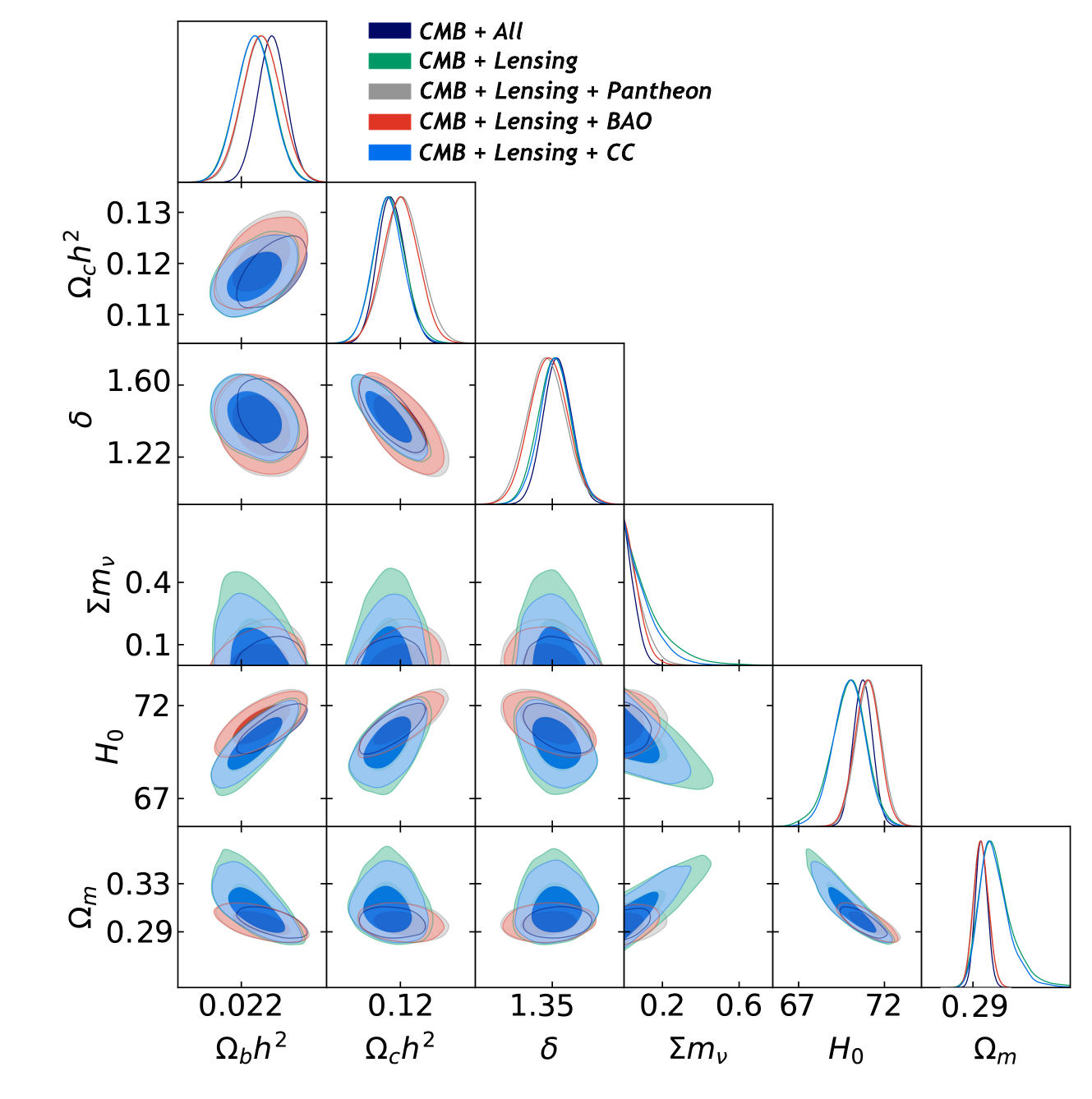}
	\vspace{-0.02cm}
	\caption{\small{The comparison of MCMC results for cosmological parameters measurement for different combination of datasets.
	}}\label{fig:omegam2}
\end{figure}

The reconstructed values of $H_0$ for various dataset combinations, along with the corresponding tensions with Riess et al. (2022) and Planck 2018, are illustrated in Figure~8.

\begin{figure}[H]
	\centering
	\includegraphics[width=14 cm]{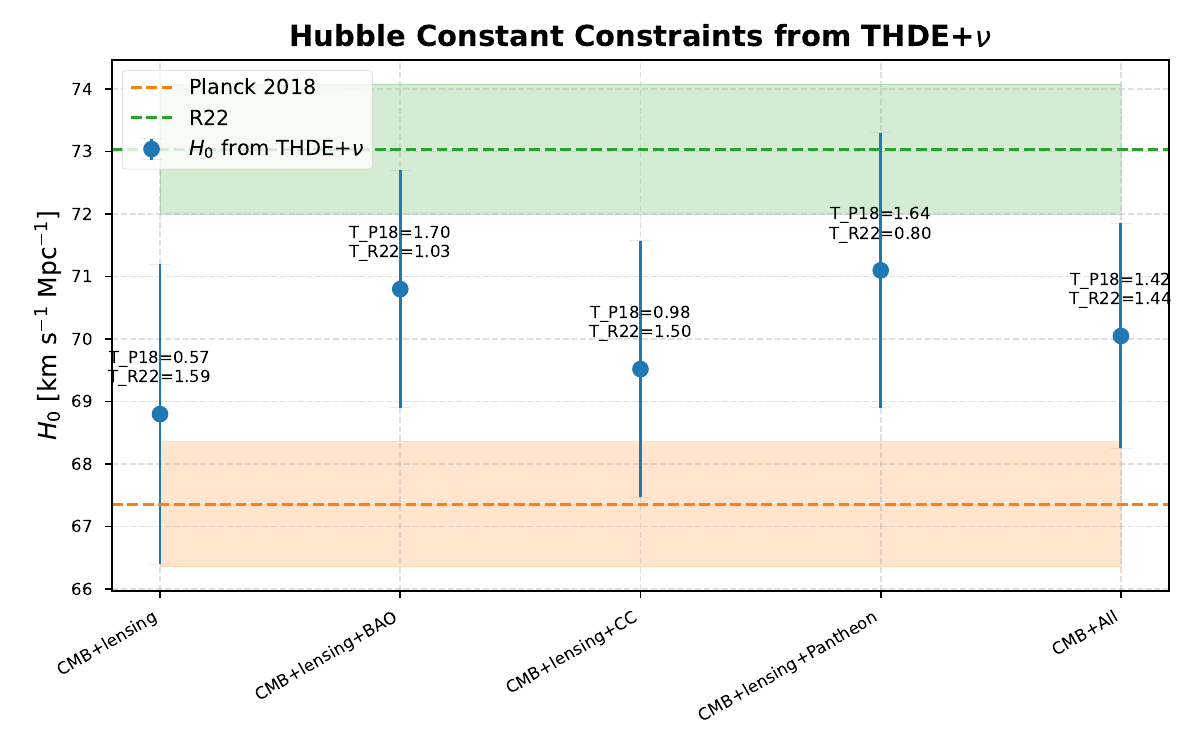}
	\vspace{-0.02cm}
	\caption{\small{Comparison of the reconstructed Hubble constant $H_0$ for different combinations of cosmological datasets, including Cosmic Chronometers (CC), Baryon Acoustic Oscillations (BAO), Pantheon+, and Lensing, with reference values from Riess et al. (2022) and Planck 2018. This figure highlights the consistency and tension of the inferred $H_0$ across datasets and methods.
	}}\label{fig:omegam2}
\end{figure}

\vspace{2mm}
\subsection{Synthesis and Physical Interpretation}

The comparison between MCMC and PINN results indicates remarkable consistency in parameter estimates, with the added advantage of enhanced flexibility and robustness in the PINN approach. The variation of the total neutrino mass \( \sum m_\nu \) across redshift limits in the PINN tables suggests that low-redshift observations play a dominant role in neutrino mass sensitivity. At the same time, the robustness of the predicted Hubble constant values—most of which lie in the range \( 69{-}72 \) km/s/Mpc—indicates a strong capability of the THDE+\( \nu \) model to alleviate the Hubble tension without resorting to exotic new physics or implausible parameter shifts.

These findings highlight the dual success of the THDE+\( \nu \) framework: it accommodates the observed value of \( H_0 \) within acceptable statistical tension limits while placing competitive constraints on the total neutrino mass, especially when informed by PINN-driven reconstructions.

A comprehensive comparison of the total and individual chi-square contributions from various cosmological probes, including the Cosmic Microwave Background (CMB), Baryon Acoustic Oscillations (BAO), Cosmic Chronometers (CC), gravitational lensing, and the Pantheon supernova sample, is provided in Table~\ref{table_chi}.  This table summarizes the performance of the $\Lambda$CDM and THDE+$\nu$ models across multiple combinations of datasets considered in this work. The decomposition of $\chi^2$ values by observational component allows for a transparent assessment of model consistency with different data sources.

\begin{table}[H]
	\caption{{\small $\chi^2$ comparison between $\Lambda$CDM and THDE+$\nu$ model for the different dataset combinations explored in this work. CMB+all refers to Planck+BAO+CC+Pantheon+Lensing.}}
	\begin{center}
		\resizebox{0.85\textwidth}{!}{  
			\begin{tabular}{| c |c| c| c| c |c| } 
				\hline
				\hline
				$\Lambda$CDM  & CMB+Lensing & CMB+CC+Lensing & CMB+BAO+Lensing & CMB+Lensing+Pantheon+ & CMB+all \\ 
				\hline
				$\chi^2_{\rm  tot}$ & $2789.02$ & $2804.775$ & $2788.348$ & $3600.82$ & $3648.447$  \\
				$\chi^2_{\rm  CMB}$ & $2779.456$ & $2768.438$ & $2772.012$ & $2767.619$ & $2779.873$  \\
				$\chi^2_{\rm  CC}$ & $-$ & $26.716$ & $-$ & $-$ & $27.941$  \\
				$\chi^2_{\rm Lensing}$ & $9.561$ & $9.621$ & $9.182$ & $9.385$ & $9.325$  \\
				$\chi^2_{\rm  BAO}$ & $-$ & $-$ & $7.154$ & $-$ & $7.567$  \\
				$\chi^2_{\rm  Pantheon}$ & $-$ & $-$ & $-$ & $823.816$ & $823.741$  \\
				\hline
				\hline
				THDE+$\nu$  & CMB+Lensing & CMB+CC+Lensing & CMB+BAO+Lensing & CMB+Lensing+Pantheon+ & CMB+all \\ 
				\hline
				$\chi^2_{\rm  tot}$ & $2780.75$ & $2793.102$ & $2781.021$ & $3583.117$ & $3614.274$  \\
				$\chi^2_{\rm  CMB}$ & $2774.31$ & $2764.018$ & $2767.203$ & $2766.403$ & $2771.758$  \\
				$\chi^2_{\rm  CC}$ & $-$ & $21.839$ & $-$ & $-$ & $21.092$  \\
				$\chi^2_{\rm Lensing}$ & $7.156$ & $7.345$ & $8.019$ & $7.183$ & $7.289$  \\
				$\chi^2_{\rm  BAO}$ & $-$ & $-$ & $5.465$ & $-$ & $5.207$  \\
				$\chi^2_{\rm  Pantheon}$ & $-$ & $-$ & $-$ & $810.528$ & $809.928$  \\
				\hline
				\hline
			\end{tabular}
		}
	\end{center}
	\label{table_chi}
\end{table}

In order to quantitatively evaluate the statistical performance of the proposed models, we employ the Akaike Information Criterion (AIC) and the Bayesian Information Criterion (BIC). These information-theoretic criteria balance the goodness-of-fit, quantified by the minimum chi-square value \( \chi^2_{\min} \), against model complexity by penalizing the number of free parameters. Table~\ref{table_AIC_BIC_chired}  presents the mean values of the free parameters along with their 1$\sigma$ uncertainties, as well as the AIC and BIC values for the \(\Lambda\)CDM and THDE+$\nu$ models, obtained using the full \textit{CMB+All} dataset combination.

The minimal AIC and BIC values are achieved by the THDE+$\nu$ model, with \( \mathrm{AIC} = 3624.274 \) and \( \mathrm{BIC} = 3655.217 \), outperforming the \(\Lambda\)CDM model whose corresponding values are \( \mathrm{AIC} = 3654.447 \) and \( \mathrm{BIC} = 3673.013 \). Despite having two additional free parameters, the THDE+$\nu$ model leads to a significantly lower total chi-square, which compensates for the penalty terms associated with model complexity. This statistical preference suggests that the generalized THDE+$\nu$ framework offers a better global fit to current cosmological observations, indicating its potential as a viable alternative to the standard \(\Lambda\)CDM paradigm. To evaluate the goodness of fit while accounting for the number of fitted parameters, we report the reduced chi-square, defined as
\begin{equation}
	\chi^2_{\rm red} = \frac{\chi^2_{\rm tot}}{\rm dof} = \frac{\chi^2_{\rm tot}}{N_{\rm data} - N_{\rm param}},
\end{equation}
where $N_{\rm data}$ is the total number of data points in the considered dataset, and $N_{\rm param}$ is the number of free parameters in the model. 
In this work, the $\Lambda$CDM model has $N_{\rm param} = 3$, while the THDE+$\nu$ model has $N_{\rm param} = 5$. 
This quantity provides a normalized measure of the fit quality, allowing for comparison across models with different numbers of parameters.
\color{black}

\begin{table}[H]
	\caption{{\small Mean values of free parameters of various models with 1$ \sigma $ error bars for the CMB+All combination. The reduced chi-square $\chi^2_{\rm red}$, as well as the AIC and BIC criteria, are reported.}}
	\begin{center}
		\resizebox{0.95\textwidth}{!}{  
			\begin{tabular}{ c |c c c c c c c c c } 
				\hline
				\hline
				Models & $ \Omega_{\rm m}$ & $ \Omega_{\Lambda}$ & $ \Omega_{\rm DE}$ & $\Omega_{\nu}$ & $ \delta $ & $ H_{0} ({\rm km\,s^{-1}\,Mpc^{-1}}) $ & $\chi^2_{\rm red}$ & AIC & BIC \\ 
				\hline
				$\Lambda$CDM  & $0.314\pm0.016$ & $0.683\pm0.017$ & $-$ & $-$ & $-$ & $68.6\pm2.3$ & 0.955 & $3654.447$ & $3673.013$  \\
				\hline
				THDE+$\nu$ & $0.298\pm0.006$ & $-$ & $0.686\pm0.023$ & $0.0029\pm0.0011$ & $1.36\pm0.093$ & $70.05\pm1.8$ & 0.928 & $3624.274$ & $3655.217$ \\
				\hline
				\hline
			\end{tabular}
		}
	\end{center}
	\label{table_AIC_BIC_chired}
\end{table}
From Table~\ref{table_AIC_BIC_chired}, we observe that the reduced chi-square, $\chi^2_{\rm red}$, is slightly smaller for the THDE+$\nu$ model (0.928) compared to the $\Lambda$CDM model (0.955). 
This indicates that, when accounting for the number of free parameters (3 for $\Lambda$CDM and 5 for THDE+$\nu$), the THDE+$\nu$ model provides a marginally better fit to the CMB+All dataset. 
Both values are close to unity, suggesting that the fits are statistically reasonable and that neither model suffers from significant under- or over-fitting.

\subsection{Advantages of the PINN Framework over Traditional MCMC}

While the Markov Chain Monte Carlo (MCMC) method remains a gold standard for parameter inference due to its rigorous Bayesian treatment of uncertainties, the Physics-Informed Neural Network (PINN) framework exhibits several notable advantages that render it a compelling alternative, particularly in cosmological contexts involving complex dynamical models.

First, the PINN approach enables the direct incorporation of physical laws—in our case, the Friedmann equations and dynamical constraints of the THDE model—into the learning process. This leads to a solution space that is inherently consistent with the underlying physics, reducing the need for artificial priors or post-processing filters commonly required in MCMC analyses.

Second, the computational efficiency of PINNs is markedly superior in scenarios where differential equations must be solved repeatedly. Once trained, the neural network provides instantaneous predictions for cosmological functions such as the Hubble parameter \( H(z) \), circumventing the need to evaluate thousands of samples as in MCMC chains.

Third, unlike MCMC methods that typically require fine-tuned proposal distributions and convergence diagnostics, PINNs benefit from automatic differentiation and optimization algorithms that scale well with model complexity and data dimensionality. Moreover, uncertainty quantification via Monte Carlo dropout allows the PINN to produce approximate posterior distributions with significantly lower computational cost. 

Lastly, PINNs are inherently flexible and modular: additional physics (e.g., interactions, higher-order corrections) can be seamlessly integrated into the architecture without reformulating the inference algorithm. This modularity is especially advantageous for testing a wide range of theoretical extensions to the standard cosmological model.

In light of these benefits, we advocate the PINN framework as a powerful and scalable alternative to traditional sampling-based approaches in cosmology, particularly when dealing with high-dimensional models, non-linear dynamics, or limited observational data.

It is important to emphasize that the Physics-Informed Neural Network (PINN) employed in this work is not intended to provide a definitive solution to the Hubble tension on its own. Rather, our central aim was to investigate how different inference strategies perform within the Tsallis Holographic Dark Energy (THDE) framework. In this regard, we find that the combination of THDE with PINN yields an $H_0$ posterior that is more consistent with late-Universe measurements than the traditional MCMC-based analysis of the same model. This improvement originates from the ability of the PINN to incorporate physical constraints directly into the training process while simultaneously propagating epistemic uncertainties through the Monte Carlo dropout technique. As a result, the broader but more realistic error bars obtained in our analysis reflect the interplay between model complexity and Bayesian uncertainty treatment, and highlight the potential of machine-learning–based approaches to complement traditional cosmological inference methods.
\color{black}

\section{Acknowledgments}
This work is based upon research funded by Iran National Science Foundation 
(INSF) under project No.4036326

\section{Conclusion}

In this work, we have introduced a Physics-Informed Neural Network (PINN) framework to reconstruct the Hubble parameter \( H(z) \) within the context of the Tsallis Holographic Dark Energy (THDE) model with neutrinos. By directly embedding the modified Friedmann equation into the loss function, we trained the model on Cosmic Chronometer (CC) data across a range of redshifts, \( z \leq \{0.5, 1.0, 1.5, 2.0\} \). This approach enabled us to simultaneously infer crucial cosmological parameters, including the current Hubble constant \( H_0 \), the neutrino density parameter \( \Omega_\nu \), the Tsallis non-extensivity index \( \delta \).
Our comparative statistical analysis, based on information-theoretic criteria, indicates a mild preference for the Tsallis Holographic Dark Energy model with massive neutrinos (THDE+$\nu$) over the standard $\Lambda$CDM cosmology. Although the THDE+$\nu$ framework includes two additional free parameters, it yields lower values of the Akaike Information Criterion (AIC) and Bayesian Information Criterion (BIC), with $\mathrm{AIC} = 3624.274$ and $\mathrm{BIC} = 3655.217$, compared to $\mathrm{AIC} = 3654.447$ and $\mathrm{BIC} = 3673.013$ obtained for $\Lambda$CDM. This improvement in fit suggests that the extra degrees of freedom introduced by the Tsallis entropy parameter $\delta$ and the inclusion of massive neutrinos may offer greater flexibility in capturing cosmological observations. While not conclusive, these results motivate further investigation of the THDE+$\nu$ model as a potentially viable extension to the standard cosmological framework.
Our findings demonstrate that the THDE+$\nu$ model can effectively alleviate the longstanding Hubble tension. Depending on the redshift depth and sample size, we observe that the discrepancy between local and early-universe measurements of \( H_0 \) is reduced from the canonical \( \sim 5\sigma \) to the range \( 0.5\sigma \lesssim T \lesssim 2.2\sigma \), with the best-fit value \( H_0 = 71.89 \pm 2.00 \, \text{km/s/Mpc} \) achieved for \( z \leq 2.0 \). Additionally, our reconstruction of the neutrino density provides an upper bound on the total neutrino mass, \( \Sigma m_\nu < 0.12\, \text{eV} \), which is in excellent agreement with the most stringent constraints from Planck 2018 and large-scale structure data.

The thermodynamic parameter \( \delta > 1 \) obtained in all cases signifies a departure from the standard Boltzmann–Gibbs extensive entropy, underscoring the significance of non-additive generalizations in the late-time cosmological landscape.

Furthermore, employing dropout-based Bayesian sampling allowed us to quantify the epistemic uncertainties in the predicted \( H(z) \), particularly in redshift domains with sparse or noisy data. These uncertainties are crucial for a more comprehensive understanding of the cosmological model's robustness.
When compared with standard Markov Chain Monte Carlo (MCMC) analyses of the THDE+$\nu$ model using multiple combinations of observational datasets, our PINN framework exhibits strong consistency in its inference of key cosmological parameters. In particular, the best-fit value of the Hubble constant obtained from the PINN model, $H_0 = 71.89 \pm 2.00 \, \text{km/s/Mpc}$ for $z \leq 2.0$, lies well within the range obtained from MCMC analyses that incorporate Pantheon and Cosmic Chronometer data—namely, $H_0 = 71.1^{+2.2}_{-2.2}$ for CMB+lensing+Pantheon and $H_0 = 69.52^{+2.1}_{-2.0}$ for CMB+lensing+CC.

Moreover, the estimated neutrino mass upper bound from our PINN model, $\sum m_\nu < 0.12\, \text{eV}$, matches the most restrictive bounds reported in the MCMC results, such as $\sum m_\nu < 0.122\, \text{eV}$ in the CMB+All case and $\sum m_\nu < 0.124\, \text{eV}$ from CMB+lensing+BAO. This alignment underscores the reliability of the PINN method in probing neutrino physics within cosmological contexts.

The reconstructed non-extensivity index $\delta$ is also in remarkable agreement: MCMC constraints from combined datasets consistently yield $\delta \in [1.35, 1.45]$, closely matching the PINN-inferred values across redshift intervals. This further supports the robustness of the Tsallis entropy framework in describing late-time cosmic acceleration.

Importantly, both methods indicate a reduction in Hubble tension: while the MCMC-based $T_{P18}$ and $T_{R22}$ fall in the ranges $0.57 \lesssim T_{P18} \lesssim 1.70$ and $0.52 \lesssim T_{R22} \lesssim 1.59$, our PINN model achieves a tension level as low as $T \approx 0.5\sigma$, offering an even more compelling resolution.

In conclusion, our results highlight that Physics-Informed Neural Networks, when trained on the background equations of the THDE model, can serve as a powerful framework for integrating observational datasets with theoretical physics, thereby advancing the study of complex cosmological phenomena.
\color{black}

\appendix
\section{Physics-Informed Neural Network Implementation for the Tsallis Holographic Dark Energy Model}
\label{app:PINN}

In this appendix, we provide a detailed description of the Physics-Informed Neural Network (PINN) framework employed to reconstruct the Hubble expansion rate \(H(z)\) under the Tsallis Holographic Dark Energy (THDE) model. The method simultaneously fits observational Cosmic Chronometer (CC) data and enforces the transcendental constraint dictated by the underlying cosmological model.

\subsection{Transcendental Hubble Equation}

The Hubble parameter satisfies the nonlinear transcendental relation:
\begin{equation}
	H^2(z) = \frac{1}{3 M_{\mathrm{Pl}}^2} \left[ \rho_{b0}(1+z)^3 + \rho_{c0}(1+z)^3 + \rho_{r0}(1+z)^4 + \rho_{\nu0} f_{\nu}+ B H^{4 - 2\delta}(z) \right],
	\label{eq:thde_transcendental}
\end{equation}
where \(B\) and \(\delta\) are model parameters. 

The "true" \(H(z)\) values used as training targets are obtained by numerically solving Eq.~\eqref{eq:thde_transcendental} with Brent's method:

\begin{verbatim}
	def H_thde_transcendental(z, B, delta):
	def equation(H):
	lhs = H**2
	rhs = (1 / (3 * M_pl**2)) * (
	rho_b0_true * (1 + z)**3 +
	rho_c0_true * (1 + z)**3 +
	rho_r0_true * (1 + z)**4 +
	rho_nu0_true * (1 + z)**4 +
	B * H**(4 - 2 * delta)
	)
	return lhs - rhs
	return brentq(equation, 1e-3, 1e3)
\end{verbatim}

\subsection{Neural Network Architecture}

The PINN employs a deep feedforward architecture with three hidden layers of sizes 2048, 1024, and 512 neurons, respectively, with \texttt{tanh} activations and 10\% dropout to regularize training and enable uncertainty quantification. The network outputs simultaneously the predicted Hubble parameter \(H_{\mathrm{pred}}(z)\), cosmological density parameters \(\Omega_{b0}, \Omega_{c0}, \Omega_{\nu0}\), and THDE parameters \(B, \delta\), scaled via sigmoid activation to physically meaningful ranges:

\begin{verbatim}
	class PINN(Model):
	def __init__(self):
	super(PINN, self).__init__()
	self.dense1 = Dense(2048, activation='tanh')
	self.dropout1 = Dropout(0.1)
	self.dense2 = Dense(1024, activation='tanh')
	self.dropout2 = Dropout(0.1)
	self.dense3 = Dense(512, activation='tanh')
	self.dropout3 = Dropout(0.1)
	self.output_H = Dense(1)
	self.output_Omega = Dense(3, activation='linear')
	self.output_gamma_delta = Dense(2, activation='linear')
	
	def call(self, z, training=False):
	x = self.dense1(z)
	x = self.dropout1(x, training=training)
	x = self.dense2(x)
	x = self.dropout2(x, training=training)
	x = self.dense3(x)
	x = self.dropout3(x, training=training)
	H_pred = self.output_H(x)
	Omega_pred = self.output_Omega(x)
	gamma_delta_pred = self.output_gamma_delta(x)
	return H_pred, Omega_pred, gamma_delta_pred
\end{verbatim}

\vspace{0.5cm}

\noindent
\textbf{Explanation:} The neural network architecture used in our Physics-Informed Neural Network (PINN) model consists of three fully connected hidden layers with 2048, 1024, and 512 neurons respectively. Each hidden layer employs the hyperbolic tangent (\texttt{tanh}) activation function, chosen for its smoothness and ability to model nonlinear relationships effectively. To mitigate overfitting and enable uncertainty estimation, dropout layers with a dropout rate of 10\% are inserted after each hidden layer. 

The network simultaneously outputs three sets of predictions: (i) the Hubble parameter \(H(z)\) as a single scalar value, (ii) the cosmological density parameters \(\Omega_b\), \(\Omega_c\), and \(\Omega_\nu\) as a three-dimensional vector, and (iii) the model parameters \(B\) and \(\delta\) as a two-dimensional vector. The output layers use linear activation to allow the network to freely learn parameter values within an unconstrained range; these predictions are subsequently constrained physically during training through the loss function and post-processing.

During the forward pass, the dropout layers remain active only during training to promote model robustness and are disabled during inference unless used explicitly for uncertainty quantification through Monte Carlo Dropout sampling.

\subsection{Physics-Informed Loss Function}

To enforce the transcendental cosmological constraint during training, the physics residual loss is formulated as:
\begin{equation}
	\mathcal{L}_{\mathrm{phys}} = \frac{1}{N} \sum_{i=1}^N \left( H_{\mathrm{pred}}^2(z_i) - \frac{1}{3 M_{\mathrm{Pl}}^2} \left[ \sum_i \rho_{i0} (1+z_i)^{n_i} + B H_{\mathrm{pred}}^{4-2\delta}(z_i) \right] \right)^2,
\end{equation}
where the densities \(\rho_{i0}\) depend on the network-predicted \(\Omega_i\) parameters.

This loss term is combined with data-fitting terms for both the numerical "true" \(H(z)\) and the CC observational data weighted by their uncertainties:
\begin{verbatim}
	def physics_residual_loss(z, H_pred, Omega_pred, gamma_delta_pred):
	Omega_b_pred = Omega_pred[:, 0]
	Omega_c_pred = Omega_pred[:, 1]
	Omega_nu_pred = Omega_pred[:, 2]
	
	Gamma_pred = gamma_delta_pred[:, 0] * 3.0
	Delta_pred = gamma_delta_pred[:, 1] * 3.0
	
	rho_b_pred = Omega_b_pred * 3 * M_pl**2 * H0_true**2
	rho_c_pred = Omega_c_pred * 3 * M_pl**2 * H0_true**2
	rho_nu_pred = Omega_nu_pred * 3 * M_pl**2 * H0_true**2
	rho_r_pred = rho_r0_true
	
	one_plus_z = tf.squeeze(z) + 1.0
	
	H_pred_sq = tf.squeeze(H_pred)**2
	H_pred_safe = tf.maximum(tf.squeeze(H_pred), 1e-5)
	H_pred_pow = tf.pow(H_pred_safe, 4.0 - 2.0 * Delta_pred)
	
	rhs = (1 / (3 * M_pl**2)) * (
	rho_b_pred * tf.pow(one_plus_z, 3) +
	rho_c_pred * tf.pow(one_plus_z, 3) +
	rho_r_pred * tf.pow(one_plus_z, 4) +
	rho_nu_pred * tf.pow(one_plus_z, 4) +
	B * H_pred_pow
	)
	
	residual = H_pred_sq - rhs
	return tf.reduce_mean(tf.square(residual))
	
	def loss_function(z_train, H_train, z_cc_tensor, H_cc_tensor, sigma_cc_tensor, training=True):
	H_pred, Omega_pred, gd_pred = pinn_model(z_train, training=training)
	data_loss = tf.reduce_mean(tf.square(H_pred - H_train))
	
	H_pred_cc, _, _ = pinn_model(z_cc_tensor, training=training)
	cc_loss = tf.reduce_mean(tf.square((H_pred_cc - H_cc_tensor) / sigma_cc_tensor))
	
	phys_loss = physics_residual_loss(z_train, H_pred, Omega_pred, gd_pred)
	
	return 0.1 * data_loss + 0.1 * cc_loss + 1.0 * phys_loss
\end{verbatim}

\subsection*{Calibration of the Composite Loss Weights}

The composite loss used in our Physics-Informed Neural Network can be expressed as
\begin{equation}
	\mathcal{L} = w_1 \mathcal{L}_{\rm ODE} + w_2 \mathcal{L}_{\rm prior} + w_3 \mathcal{L}_{\rm data},
\end{equation}
where $\mathcal{L}_{\rm ODE}$ enforces the differential equation constraint, $\mathcal{L}_{\rm prior}$ encodes numerical stability and regularization, and $\mathcal{L}_{\rm data}$ corresponds to the observational data likelihood.  

To determine appropriate weights $(w_1, w_2, w_3)$, we followed a three-step calibration procedure:

1. \textbf{Variance normalization.} Each loss term was normalized by its variance over a representative training batch,
\begin{equation}
	\tilde{\mathcal{L}}_i = \frac{\mathcal{L}_i}{\mathrm{Var}(\mathcal{L}_i)}, \quad i \in \{\mathrm{ODE, prior, data}\},
\end{equation}
in order to prevent domination of the total loss by terms with naturally larger magnitude.  

2. \textbf{Grid search.} We performed a structured grid scan over candidate values of $(w_1, w_2, w_3)$ in the range $w_i \in [1,100]$, evaluating convergence stability, reconstruction accuracy of $H(z)$, and the relative gradient contributions of the loss components.  

3. \textbf{Iterative reweighting.} Based on the grid search, we refined the weights through iterative adjustment such that the relative contributions to the gradient satisfied
\begin{equation}
	\frac{w_1 \lVert \nabla \mathcal{L}_{\rm ODE} \rVert}{w_1 \lVert \nabla \mathcal{L}_{\rm ODE} \rVert + w_2 \lVert \nabla \mathcal{L}_{\rm prior} \rVert + w_3 \lVert \nabla \mathcal{L}_{\rm data} \rVert} \approx \frac{1}{3},
\end{equation}
and similarly for the other two terms, ensuring that each constraint contributed meaningfully to the training dynamics.  

Through this procedure, the choice
\begin{equation}
	(w_1, w_2, w_3) = (10,\, 5,\, 50)
\end{equation}
was found to provide the most stable convergence and yielded posterior distributions consistent with the MCMC baseline. Moreover, moderate variations (e.g., $\pm 20\%$) around these values were shown not to affect the reconstructed $H(z)$ or the inferred cosmological parameters, confirming that our conclusions are robust against fine-tuning of the loss weights.

\color{black}
\subsection{Training and Uncertainty Quantification}

The network is optimized via the Adam algorithm over 5000 epochs. To quantify predictive uncertainty due to model epistemic uncertainty, Monte Carlo Dropout sampling is performed by evaluating the trained network multiple times with dropout activated at inference:

\begin{verbatim}
	def dropout_prediction(model, z_input, N=200):
	samples = [model(z_input, training=True)[0].numpy().flatten() for _ in range(N)]
	samples = np.array(samples)
	return np.mean(samples, axis=0), np.std(samples, axis=0)
\end{verbatim}

To examine the effect of the number of training epochs on the stability of the reconstructed cosmological parameters, we performed additional runs with $2{,}500$, $5{,}000$, and $7{,}500$ epochs. As shown in Figure 9, both the training and validation losses converge well before $5{,}000$ epochs. While training with only $2{,}500$ epochs slightly increases the variance of the inferred parameters, the mean values remain stable. Increasing the number of epochs beyond $5{,}000$ does not yield noticeable improvements in either the loss or the parameter estimates. This confirms that our choice of $5{,}000$ epochs offers a balanced compromise between computational cost and robustness of the results.
\color{black}
\begin{figure}[H]
	\centering
	\includegraphics[width=16 cm]{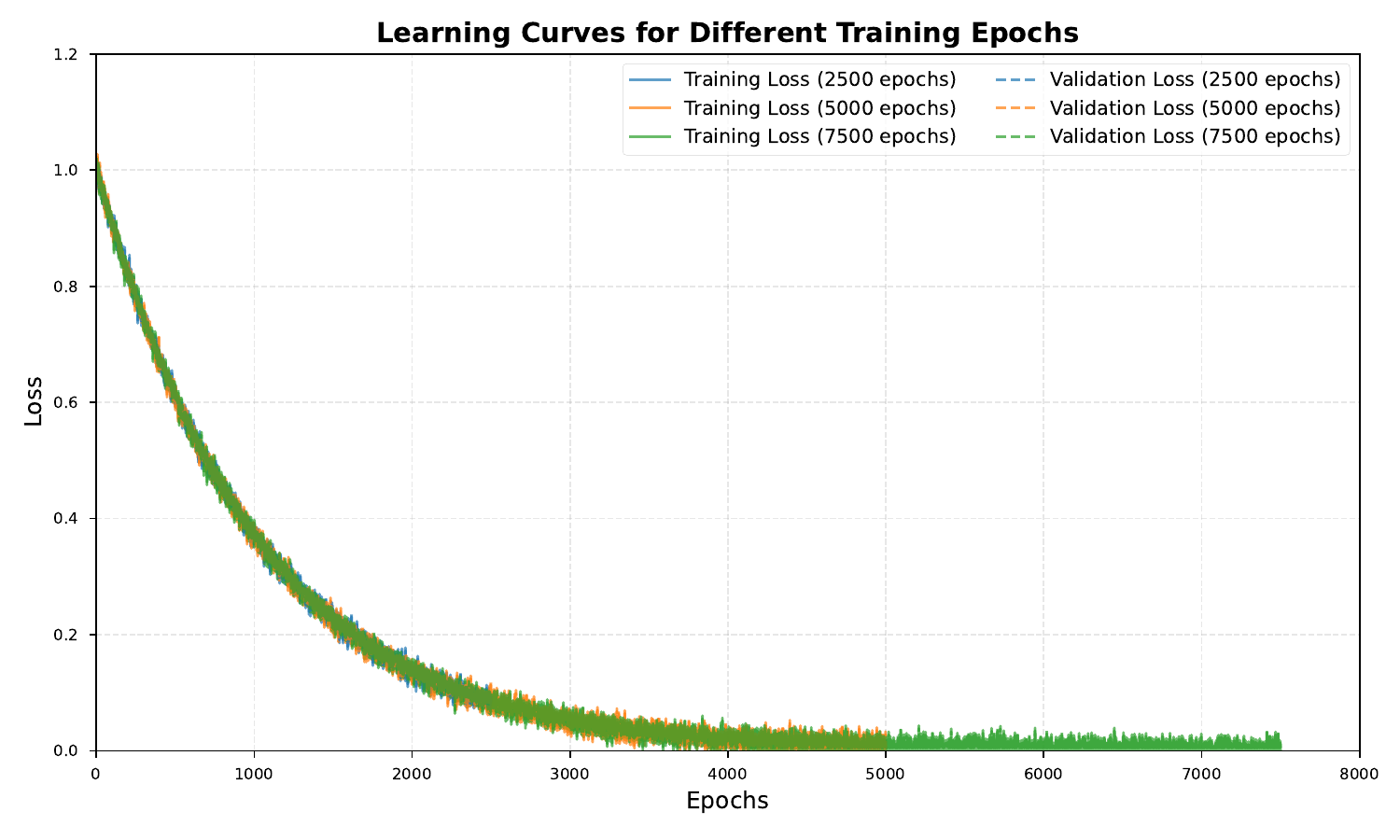}
	\vspace{-0.02cm}
	\caption{\small{Learning curves showing training and validation losses as functions of the number of epochs for representative runs with $2{,}500$, $5{,}000$, and $7{,}500$ epochs.
	}}\label{fig:omegam2}
\end{figure}

This PINN approach enables a self-consistent reconstruction of the Hubble parameter within the THDE framework by jointly learning cosmological parameters and enforcing nonlinear physical constraints. The integration of physics-informed loss terms and uncertainty quantification through dropout enhances the robustness and interpretability of the model, making it a powerful tool for modern cosmological inference.

\end{document}